\pdfoutput=1

\documentclass[11pt,twoside,a4paper,cmspaper,final,collab]{cms-tdr}
\def\svnVersion{aa8e976-D}\def\svnDate{2020/02/09}\def\cmsCernNoTag{CERN-EP-2019-257}\def\cmsCernDate{\today}\def\cmsMessage{"Published in the Journal of High Energy Physics as \href{http://dx.doi.org/10.1007/JHEP03(2020)131}{\doi{10.1007/JHEP03(2020)131}.}"}

\begin{document}\cmsNoteHeader{HIG-18-031}

\hyphenation{had-ron-i-za-tion}
\hyphenation{cal-or-i-me-ter}
\hyphenation{de-vices}
\newcommand{\mh}{\ensuremath{m_{\PH}}}
\newcommand{\topq}{\PQt}
\newcommand{\bq}{\PQb}
\newcommand{\cq}{\PQc}
\newcommand{\cqbar}{\PAQc}
\newcommand{\xmark}{\ensuremath{\checkmark}}
\newcommand{\vhcc}{\ensuremath{\PV\PH(\PH\to \PQc \PAQc)}\xspace}
\newcommand{\vhtautau}{\ensuremath{\PV\PH(\PH\to \tau \tau)}\xspace}
\newcommand{\zhcc}{\ensuremath{\PZ\PH(\PH\to \PQc \PAQc)}\xspace}
\newcommand{\whcc}{\ensuremath{\PW\PH(\PH\to \PQc \PAQc)}\xspace}
\newcommand{\vhbb}{\ensuremath{\PV\PH(\PH\to \cPqb \cPaqb)}\xspace}
\newcommand{\vzcc}{\ensuremath{\PV\PZ(\PZ\to \PQc \PAQc)}\xspace}
\newcommand{\ZnnH}{\ensuremath{\PZ(\PGn\PGn)\PH}\xspace}
\newcommand{\ZmmH}{\ensuremath{\PZ(\PGm\PGm)\PH}\xspace}
\newcommand{\ZeeH}{\ensuremath{\PZ(\Pe\Pe)\PH}\xspace}
\newcommand{\WlnHcc}{\ensuremath{\PW(\ell\PGn)\PH(\ccbar)}\xspace}
\newcommand{\ZllHcc}{\ensuremath{\PZ(\ell\ell)\PH(\ccbar)}\xspace}
\newcommand{\ZnnHcc}{\ensuremath{\PZ(\PGn\PGn)\PH(\ccbar)}\xspace}
\newcommand{\met}{\ptmiss}
\newcommand{\metvec}{\ptvecmiss}
\newcommand{\WmnH}{\ensuremath{\PW(\PGm\PGn)\PH}\xspace}
\newcommand{\WenH}{\ensuremath{\PW(\Pe\PGn)\PH}\xspace}
\newcommand{\HCC}{\ensuremath{\PH\to\ccbar}\xspace}
\newcommand{\HBB}{\ensuremath{\PH\to\bbbar}\xspace}
\newcommand{\Vpt}{\ensuremath{\pt(\PV)}}
\newcommand{\Vudsg}{\PV{}+\PQu{}\cPqd{}\cPqs{}\Pg}
\newcommand{\Vbb}{\PV{}+\PQb{}\cPaqb{}}
\newcommand{\Vcc}{\PV{}+\PQc{}\PAQc{}}
\newcommand{\pb}{\unit{pb}}
\newcommand{\zh}{\PZ{}\PH\xspace}
\newcommand{\vh}{\PV\PH\xspace}
\newcommand{\Wmn}{\ensuremath{\PW(\mu\nu)}}
\newcommand{\Zmm}{\ensuremath{\PZ(\PGm\PGm})}
\newcommand{\Zee}{\ensuremath{\PZ(\Pe\Pe)}}
\newcommand{\dphiVH}{\ensuremath{\Delta\phi(\PV,\PH)}}
\newcommand{\ptV}{\ensuremath{\pt(\PV)}\xspace}
\newcommand{\ptH}{\ensuremath{\pt(\PH)}\xspace}
\newcommand{\larger}{large-$R$}
\newcommand{\Larger}{Large-$R$}
\newcommand{\smallr}{small-$R$}
\newcommand{\nsmallr}{\ensuremath{\text{N}^{\text{aj}}_{\text{small-}R}}}
\newcommand{\msd}{\ensuremath{m_{\text{SD}}}}
\newcommand{\VZ}{\ensuremath{\PV\PZ}\xspace}
\newcommand{\VH}{\ensuremath{\PV\PH}\xspace}
\newcommand{\WH}{\ensuremath{\PW\PH}\xspace}
\newcommand{\ZH}{\ensuremath{\PZ\PH}\xspace}
\newcommand{\WW}{\ensuremath{\PW\PW}\xspace}
\newcommand{\WZ}{\ensuremath{\PW\PZ}\xspace}
\newcommand{\ZZ}{\ensuremath{\PZ\PZ}\xspace}
\newcommand{\jone}{\ensuremath{\text{j}_1}\xspace}
\newcommand{\jtwo}{\ensuremath{\text{j}_2}\xspace}

\cmsNoteHeader{HIG-18-031}

\title{A search for the standard model Higgs boson decaying to charm quarks}
\date{\today}

\abstract{
A direct search for the standard model Higgs boson, \PH, produced in association with a vector boson, \PV\ (\PW or \PZ), and decaying to a charm quark pair is presented. The search uses a data set of proton-proton collisions corresponding to an integrated luminosity of 35.9\fbinv, collected by the CMS experiment at the LHC in 2016, at a centre-of-mass energy of 13\TeV. The search is carried out in mutually exclusive channels targeting specific decays of the vector bosons: $\PW\to \ell \nu$, $\PZ\to \ell\ell$, and $\PZ\to \nu\nu$, where $\ell$ is an electron or a muon. To fully exploit the topology of the \PH boson decay, two strategies are followed. In the first one, targeting lower vector boson transverse momentum, the \PH boson candidate is reconstructed via two resolved jets arising from the two charm quarks from the \PH boson decay. A second strategy identifies the case where the two charm quark jets from the \PH boson decay merge to form a single jet, which generally only occurs when the vector boson has higher transverse momentum. Both strategies make use of novel methods for charm jet identification, while jet substructure techniques are also exploited to suppress the background in the merged-jet topology. The two analyses are combined to yield a 95\% confidence level observed (expected) upper limit on the cross section $\sigma\left(\VH\right)\mathcal{B}\left(\PH \to \ccbar \right)$ of 4.5 ($2.4^{+1.0}_{-0.7}$)\pb, corresponding to 70 (37) times the standard model prediction. }

\hypersetup{
pdfauthor={CMS Collaboration},
pdftitle={A search for the standard model Higgs boson decaying to charm quarks},
pdfsubject={CMS},
pdfkeywords={CMS, physics, Higgs boson, charm quark}}

\maketitle

\section{Introduction}
The discovery of a Higgs boson, \PH, with the CERN LHC data collected in 2010--2012 by both the ATLAS~\cite{Aad:2012tfa} and CMS~\cite{Chatrchyan:2012xdj,Chatrchyan2013} experiments in 2012
represented a major step toward the characterisation of the electroweak symmetry breaking mechanism~\cite{PhysRevLett.13.321,PhysRevLett.13.508,PhysRevLett.13.585}.
The mass of this particle is measured to be $\mh \sim 125$\GeV~\cite{Aad:2015zhl,Sirunyan:2017exp,Aaboud:2019lxo} and its decays in the $\PGg\PGg$, $\PZ\PZ$, $\PW\PW$, and $\PGt\PGt$ modes have
been observed~\cite{Aad:2014eha,Khachatryan:1728107,Aad:2014eva,Chatrchyan:2013mxa,ATLAS:2014aga,Aad:2015ona,Chatrchyan:1633401,Aad:2015vsa,Chatrchyan:1643937,Sirunyan:2273798,Sirunyan:2018egh}.
All measured properties so far~\cite{Sirunyan:2017exp,Sirunyan:2019twz,Sirunyan:2018koj,Aad:2015gba,Khachatryan:2014jba,Chatrchyan:2012jja,Aad:2013xqa,Khachatryan:2016vau,Aad:2015zhl,Sirunyan:2017exp,Aaboud:2018ezd,Aaboud:2018wps}
indicate that, within the measurement uncertainties, this new particle is consistent with the expectations of the standard model (SM).
Nevertheless, there remains much to be learned about the properties of this new particle. One of the highest priorities of the LHC physics
program is the measurement of the couplings of the \PH boson to other SM particles. Recently both ATLAS and CMS Collaborations reported the
first direct measurements of the \PH boson couplings to third-generation quarks (\topq\ and \bq )~\cite{Sirunyan:2018hoz,Aaboud:2018urx,Sirunyan:2018kst,Aaboud:2018zhk}
and found them to be also compatible with the SM prediction.
A measurement of the couplings of the \PH boson to second generation leptons~\cite{ATLAS-CONF-2019-028,Sirunyan:2018hbu} and quarks is the next target.

In this paper, we focus on the search for \PH bosons decaying to $\PQc\PAQc$, a charm quark-antiquark pair. The \PH boson to charm quark
Yukawa coupling $y_{\cq}$ can be significantly modified by physics beyond the SM~\cite{Delaunay:2013pja,Ghosh:2015gpa,Botella:2016krk,Harnik:2012pb,Altmannshofer:2016zrn}.
In the absence of an observation of Higgs decays to charm quarks, one can place a bound on the charm quark Yukawa coupling.
The first direct bound on $\kappa_{\text{c}} \equiv y_{\text{\cq}} / y_{\text{\cq}}^{\text{SM}}$ of 234 at 95\% confidence level (\CL) has been obtained by recasting the ATLAS and CMS
$8$\TeV \HBB searches~\cite{Perez:2015aoa} in a model independent way. Indirect constraints on $y_{\cq}$ obtained from a global fit to existing \PH boson data
result in an upper bound on $\kappa_{\PQc} \equiv y_{\PQc} / y_{\PQc}^{\text{SM}}$ of 6.2~\cite{Perez:2015aoa} at 95\% confidence level (\CL), assuming the absence of non-SM production mechanisms.
A direct measurement of this process is extremely challenging at a hadron collider. The branching fraction of this process according to SM
computations, $\mathcal{B}(\HCC)=0.0288^{+0.0016}_{-0.0006}$~\cite{deFlorian:2016spz}, is a factor 20 smaller than that of
$\PH \to \bbbar$, and there is a very large background from SM processes comprised uniquely of
jets produced through the strong interaction, referred to as quantum chromodynamics (QCD) multijet events.
Results from direct searches for $\HCC$ at the LHC in the \zh ($\PZ\to\ell\ell$, $\ell=\Pe$ or $\mu$) channel
were previously reported by the ATLAS Collaboration using a data sample of proton-proton ($\Pp\Pp$) collisions at a centre-of-mass energy of $13\TeV$,
corresponding to an integrated luminosity of 36.1\fbinv ~\cite{Aaboud:2018fhh}.
The observed (expected) exclusion limit on the signal strength $\mu$ (defined as the product of the measured \PH boson production cross section
and the $\PH \to \ccbar$ branching fraction divided by the same quantity as predicted by the SM) at 95\% CL was found to be 110 (150).

This paper presents the first direct search for the \HCC decay carried out by the CMS Collaboration. It uses $\Pp\Pp$ collision data corresponding to an integrated luminosity of 35.9\fbinv, collected with the CMS experiment at the LHC in 2016 at a centre-of-mass energy of 13\TeV. The search targets \PH bosons produced in association with a \PW\ or \PZ\ boson, which we collectively refer to as vector (\PV) bosons. The presence of a \PV boson greatly suppresses backgrounds stemming from otherwise overwhelming QCD multijet processes, and its leptonic decays provide a crucial handle to collect the events efficiently.
The most significant remaining backgrounds arise from \PV{}+jets (processes that account for one or more jets recoiling against a vector boson), \ttbar, and \vhbb processes.
To fully explore the \HCC decay mode, the analysis is split into two separate searches involving different topologies: the ``resolved-jet'' topology, in which the \PH boson candidate is reconstructed from two well-separated and individually resolved charm quark jets, and the ``merged-jet'' topology, in which the hadronisation products of the two charm quarks are reconstructed as a single jet. The former focuses on \PH boson candidates with lower transverse momentum, \pt, while the latter performs better for \PH boson candidates with high \pt. In practice, the two topologies can have significant overlap and so, for the final result, the two are made distinct by defining them in reference to whether the \PV boson in the event has \ptV below or above a single threshold chosen to maximise the sensitivity to the \vhcc process.

The central feature of this search is the identification of charm quark jets. In both topologies, novel tools based upon advanced machine learning (ML) techniques are used for charm quark jet identification~\cite{Sirunyan:2017ezt,CMS-PAS-JME-18-002}. In addition, the merged-jet topology makes use of jet substructure information to further suppress the backgrounds.

\section{The CMS detector}

The central feature of the CMS apparatus is a superconducting solenoid of 6\unit{m} internal diameter, providing a magnetic field of 3.8\unit{T}. Within the solenoid volume are a silicon pixel and strip tracker, a lead tungstate crystal electromagnetic calorimeter (ECAL), and a brass and scintillator hadron calorimeter (HCAL), each composed of a barrel and two endcap sections. Forward calorimeters extend the pseudorapidity ($\eta$) coverage provided by the barrel and endcap detectors. Muons are detected in gas-ionisation chambers embedded in the steel flux-return yoke outside the solenoid.

In the barrel section of the ECAL, an energy resolution of about 1\% is achieved for unconverted or late-converting photons that have energies in the range of tens of GeV. The remaining barrel photons have a resolution of about 1.3\% up to $\abs{\eta} = 1$, rising to about 2.5\% at $\abs{\eta} = 1.4$. In the endcaps, the resolution of unconverted or late-converting photons is about 2.5\%, while other endcap photons have a resolution between 3 and 4\%~\cite{CMS:EGM-14-001}.

In the region $\abs{\eta} < 1.74$, the HCAL cells have widths of 0.087 in $\eta$ and 0.087\unit{rad} in azimuth ($\phi$). In the $\eta$-$\phi$ plane, and for $\abs{\eta} < 1.48$, the HCAL cells map on to $5{\times}5$ arrays of ECAL crystals to form calorimeter towers projecting radially outwards from close to the nominal interaction point. For $\abs{\eta} > 1.74$, the coverage of the towers increases progressively to a maximum of 0.174 in $\Delta \eta$ and $\Delta \phi$. Within each tower, the energy deposits in ECAL and HCAL cells are summed to define the calorimeter tower energies.

Muons are measured in the range $\abs{\eta} < 2.4$, with detection planes made using three technologies: drift tubes, cathode strip chambers, and resistive-plate chambers. The efficiency to reconstruct and identify muons is greater than 96\%. Matching muons to tracks measured in the silicon tracker results in a relative \pt resolution, for muons with \pt up to 100\GeV, of 1\% in the barrel and 3\% in the endcaps. The \pt resolution in the barrel is better than 7\% for muons with \pt up to 1\TeV~\cite{Sirunyan:2018}.

Events of interest are selected using a two-tiered trigger system~\cite{Khachatryan:2016bia}. The first level, composed of custom hardware processors, uses information from the calorimeters and muon detectors to select events at a rate of around 100\unit{kHz} within a fixed time interval of less than 4\mus. The second level, known as the high-level trigger (HLT), consists of a farm of processors running a version of the full event reconstruction software optimised for fast processing, and reduces the event rate to around 1\unit{kHz} before data storage.

A more detailed description of the CMS detector, together with a definition of the coordinate system used and the relevant kinematic variables, can be found in Ref.~\cite{Chatrchyan:2008zzk}.

\section{Simulated event samples}
\label{sec:mcsamples}

Signal and background processes are simulated using various event
generators, while the CMS detector response is modelled with
\GEANTfour~\cite{GEANT4}. The quark-induced \ZH and \WH signal
processes are generated at next-to-leading order (NLO) accuracy in QCD
using the {\POWHEG} v2~\cite{Nason:2004rx,POWHEG,Alioli:2010xd} event
generator extended with the Multi-scale improved NLO (MiNLO)
procedure~\cite{Hamilton2012,Luisoni:2013kna}, while the gluon-induced
\ZH process is generated at leading order (LO) accuracy with {\POWHEG}
v2. The \PH boson mass is set to 125\GeV for all signal
samples. The production cross sections of the signal processes~\cite{deFlorian:2016spz} are corrected as a function of \ptV to next-to-next-to-leading order (NNLO) QCD + NLO electroweak (EW)
accuracy combining the \textsc{vhnnlo}~\cite{Ferrera:2013yga,Ferrera:2014lca,Ferrera:2011bk,Ferrera:2017zex},
\textsc{vh@nnlo}~\cite{Brein:2012ne,Harlander:2013mla}, and \textsc{hawk} v2.0~\cite{Denner:2014cla} generators
as described in Ref.~\cite{deFlorian:2016spz}.

The \PV{}+jets events are generated with {\MGvATNLO} v2.4.2~\cite{Alwall:2014hca} at NLO with up to two additional partons,
and at LO accuracy with up to four additional partons. The production cross sections for
the \PV{}+jets samples are scaled to the NNLO cross sections obtained using \FEWZ 3.1~\cite{Li:2012wna}. Events in both LO and NLO samples are reweighted
to account for NLO EW corrections to \ptV, which reach up to 10\% for $\ptV \approx 400\GeV$. In addition, a LO-to-NLO correction is applied
to LO samples as a function of the separation in $\eta$ between the two leading jets in the event~\cite{Sirunyan:2017elk}.
The \ptV spectrum in simulation after the aforementioned corrections is observed to be harder than in data, as expected due
to missing higher-order EW and QCD contributions to the \PV{}+jets processes~\cite{NLO_plus_QCD_weight}.
A residual reweighting of \ptV, that is obtained via a fit to the data-to-simulation ratio in the control regions (detailed in Section~\ref{sec:resolved_an})
of the \WlnHcc and \ZllHcc channels in the resolved analysis, is applied.

Diboson (\WW,  \WZ\ and \ZZ) background events are generated with {\MGvATNLO}
v2.4.2~\cite{Alwall:2014hca} at NLO with up to two additional partons in the matrix element calculations.
The same generator is used at LO accuracy to generate a sample of QCD multijet events. The
\ttbar~\cite{Frixione:2007nw} and single top production processes
in the $\PQt\PW$- and $t$-channels~\cite{Re:2010bp,Frederix:2012dh}
are generated to NLO accuracy with {\POWHEG} v2, while the $s$-channel~\cite{Alioli:2009je}
single top process is generated with {\MGvATNLO} v2.4.2.
The production cross sections for the \ttbar samples are scaled to the
NNLO prediction with the next-to-next-to-leading-log result obtained from {{\textsc{Top++}     v2.0}}~\cite{Czakon:2011xx}.
The \ttbar samples are reweighted as a function of top quark \pt to account for the known differences between data and simulation~\cite{PhysRevD.95.092001}.

The parton distribution functions (PDF) used to produce
all samples are the NNLO {NNPDF3.1}
set~\cite{Ball:2017nwa}. For parton showering and hadronisation, including the \HCC decay, the
matrix element generators are interfaced with {\PYTHIA}
v8.230~\cite{Sjostrand:2014zea} with the CUETP8M1~\cite{Khachatryan:2015pea} underlying event tune.
The matching of jets from matrix element calculations and those from parton shower
is done with the FxFx~\cite{Frederix:2012ps} (MLM~\cite{Alwall:2007fs}) prescription for NLO (LO) samples.
For all samples, simulated additional $\Pp\Pp$ interactions in the same or adjacent bunch crossings (pileup) are added to the hard-scattering
process. The events are then reweighted to match the pileup profile observed in the collected data.

\section{Event reconstruction and selection}
\label{sec:reconstruction}
Events are reconstructed using the CMS particle-flow (PF) algorithm~\cite{CMS-PRF-14-001},
which seeks to reconstruct and identify the individual particles in the event via an optimal combination
of all information in the CMS detector. The reconstructed particles are identified as charged or neutral hadrons,
electrons, muons, or photons, and constitute a list of PF candidate physics objects.
At least one reconstructed vertex is required. In the case of multiple collision vertices
from pileup interactions, the candidate vertex with the largest value of summed physics-object $\pt^2$ is taken to be the primary $\Pp\Pp$ interaction vertex.
The physics objects are the jets, clustered using the jet finding algorithm~\cite{Cacciari:2008gp,Cacciari:2011ma} with the tracks assigned to candidate vertices as inputs,
and the associated missing transverse momentum, taken as the negative vector sum of the \pt of those jets.
Events affected by reconstruction failures, detector malfunctions, or noncollision backgrounds,
are identified and rejected by dedicated filters~\cite{Sirunyan:2019kia}.

Electrons are reconstructed by combining information from the tracker and energy deposits
in the ECAL~\cite{Khachatryan:2015hwa}. Muons are reconstructed by combining information
from the tracker and the muon system~\cite{Sirunyan:2018}. Only tracks originating from the PV can be
associated with the electrons or muons, and quality criteria~\cite{Khachatryan:2015hwa,Sirunyan:2018}
are further imposed that obtain more pure identification without substantial loss of efficiency. To suppress leptons
stemming from \bq\ and \cq\ decays, while retaining leptons from \PV\ decays, isolation is required
from jet activity within a cone of radius $\Delta R = \sqrt{\smash[b]{(\Delta\eta)^2+(\Delta\phi)^2}} = 0.3$.
The isolation is defined as the scalar \pt\ sum of the PF candidates within the cone divided
by the lepton \pt. The upper threshold applied on the relative isolation is 0.06 for electrons and muons in the \WlnHcc channel and 0.15 and 0.25
for electrons and muons respectively in the \ZllHcc channel. Charged PF candidates not originating from the PV, as well as PF candidates
identified as electrons or muons, are not considered in the sum~\cite{CMS:2019kuk}.
The isolation of electrons and muons is also corrected for the estimated energy that is contributed to the isolation region by neutral particles originating from pileup.
In the case of electrons, the latter is estimated by an effective jet area from the measured neutral energy density~\cite{Khachatryan:2015hwa}, while for muons,
the $\Delta\beta$-correction method~\cite{Sirunyan:2018} is applied.

Jets are reconstructed by clustering the PF candidates with the anti-\kt algorithm~\cite{Cacciari:2008gp, Cacciari:2011ma}
using a distance parameter $R$. The jet momentum is determined as the vectorial sum of all PF candidate
momenta in the jet, and is found in simulation to be within 5 to 10\% of the true momentum over the full detector acceptance and range of \pt considered in this analysis. The raw jet energies are then corrected to establish a uniform response of the calorimeter in $\eta$ and a calibrated absolute response in \pt.
Additional corrections to account for any residual differences between the jet energy scale in data and simulation are extracted and applied based on comparison of data and simulated samples in relevant control regions~\cite{Khachatryan:2016kdb}.
The jet energy resolution typically amounts to 15--20\% at 30\GeV, about 10\% at 100\GeV, and 5\% at 1\TeV~\cite{Khachatryan:2016kdb}. Corrections extracted from data control regions are applied to account for the difference between the jet energy resolution in data and simulation. Additional selection criteria are applied to each jet to remove those that are potentially dominated by instrumental or reconstruction failures~\cite{CMS-PAS-JME-16-003}.

Two collections of jets reconstructed with the anti-\kt algorithm are used in the search. The first consists of jets clustered
with $R=0.4$, and will be referred to as ``\smallr\ jets''. The charged hadron subtraction algorithm~\cite{CMS-PAS-JME-14-001}
is used to eliminate PF candidates from the jet constituents associated with vertices from pileup interactions.
The neutral component of the energy arising from pileup interactions is estimated with the effective area method~\cite{CMS-PAS-JME-16-003}.
The \smallr\ jets are required to have $\pt > 20$\GeV and to be within the tracker acceptance, $\abs{\eta} < 2.4$. Any \smallr\ jets that overlap with preselected
electrons and muons, as defined by $\Delta R(\text{j},\ell)<0.4$, are discarded.

The second jet collection is based on  jets reconstructed using $R=1.5$.
This collection will be referred to as ``\larger\ jets'' in what follows.
In this case, the PUPPI algorithm~\cite{Bertolini:2014bba} is used to correct the jet energy for contributions coming from pileup.
Additional information on jet substructure is obtained by reclustering the constituents of
these jets via the Cambridge--Aachen algorithm~\cite{Dokshitzer:1997in}.
The ``modified mass drop tagger'' algorithm~\cite{Dasgupta:2013ihk,Butterworth:2008iy}, also known as the
``soft-drop'' (SD) algorithm, with angular exponent $\beta = 0$, soft cutoff threshold $z_{\text{cut}} =
0.1$, and characteristic radius $R_{0} = 1.5$~\cite{Larkoski:2014wba}, is applied to remove soft, wide-angle
radiation from the jet.
In the default configuration, the SD algorithm identifies two hard subjets within the \larger\ jet by
reversing the Cambridge--Aachen clustering history. The kinematic variables of the two subjets are used to
calculate the 4-momentum of the \larger\ jet. The \larger\ jets are required to
have $\abs{\eta}<2.4$ and a soft drop mass of $50 < \msd < 200\GeV$.
\Larger\ jets that overlap with preselected electrons and muons, as defined by $\Delta R(\text{j},\ell)<1.5$, are discarded.

The missing transverse momentum vector \ptvecmiss is computed as the negative vector \pt\ sum of all the PF candidates in an event,
and its magnitude is denoted \ptmiss~\cite{Sirunyan:2019kia}. The magnitude and direction of \ptvecmiss are modified to account for corrections to the energy
scale of the reconstructed jets in the event.

One of the most challenging tasks of this analysis is the discrimination of jets that are the result of the hadronisation
of \PQc quarks from all other jet flavours. Tagging \PQc jets is more difficult than tagging \PQb jets because
they are less distinct from light-flavour quark or gluon jets (\cPqu{}\cPqd{}\cPqs{}\Pg{}) in regard to mass, decay length of charmed hadrons
produced in the hadronisation process, and multiplicity of tracks inside the jet. The resolved- and merged-jet topology analyses use different strategies for tagging \PQc jets.
More details on \PQc tagging are presented below in Sections \ref{sec:resolvedanalysis} and \ref{sec:boostedanalysis}.

\subsection{Baseline selection}
\label{sec:selection}
The search uses the leptonic decays of the vector bosons to define three mutually exclusive channels
based on the charged-lepton multiplicity in the final state, namely: ``0L'' channel as referring to the \ZnnHcc
signal process, ``1L'' channel as referring to the \WlnHcc signal process, and ``2L'' channel as referring to the \ZllHcc signal process.
The 1L and 2L channels are further subdivided based on lepton flavour. Only electrons and muons are considered in this search.

Events in the 0L channel are collected with a trigger requiring the presence of \ptmiss\ above $170$\GeV or $110$\GeV and an additional threshold on the missing hadronic transverse energy of $110$\GeV.
Events in the 1L channel are obtained with a trigger requiring the presence of an isolated electron or muon with \pt above 27 and 24\GeV, respectively.
Events in the 2L channel of the resolved-jet topology analysis are selected by triggers that require the presence of a
pair of leptons with \pt larger than 23 and 12\GeV for electrons, and 17 and 8\GeV for muons. The same dielectron trigger has been used in the 2L \Zee\ channel of the merged-jet topology analysis,
while events in the \Zmm\ channel are selected by the above single-muon trigger, which provides high efficiency for muons produced in the decays of high-\pt bosons.

The collected events are required to pass additional offline criteria. In the 0L channel corresponding to \PZ boson decays to neutrinos, $\ptmiss > 170$\GeV
is required and events with identified isolated leptons are rejected. The \ptvecmiss\ is taken to correspond to $\ptvec(\PV)$ in this
case. Events with a single electron (muon) with $\pt>30$ (25)\GeV pass the 1L selection. The leptonically decaying
\PW boson is approximately reconstructed as the vectorial sum of the lepton momentum and \ptvecmiss.
The event topology is required to be compatible with the leptonic decay of a Lorentz-boosted \PW boson
by requiring $\Delta\phi(\ptmiss,\ell)<2.0$ (1.5) in the resolved-jet (merged-jet) topology analysis.
Finally, for the 2L selection, the two highest \pt leptons are required to be of the same flavour, opposite
electric charge, and to have a \pt\ above 20\GeV. The \PZ boson candidates are then reconstructed as the sum of the four-momenta of these two leptons,
and the invariant mass of the candidates is required to be compatible with the \PZ boson mass ($75 <m_{\ell\ell}< 105\GeV$).

A typical \vhcc\ event has the signature of a vector boson recoiling against a \PH boson with little additional activity.
The event selection is designed to retain such events while suppressing background processes as much as possible.
In addition to the requirement of a high-\pt\ vector boson,
the QCD multijet background is reduced to negligible levels by demanding the $\ptvecmiss$ to not be aligned with any jet in the event and requiring the azimuthal angular separation
$\Delta\phi(\ptvecmiss{}_{\text{trk}},\ptvecmiss)<0.5$ for which $\ptvecmiss{}_{\text{trk}}$ is calculated solely from charged particles.
This latter selection reduces the contribution of QCD multijet events that arise from the presence of "fake" $\ptvecmiss$ coming from jet energy mismeasurement in the calorimeters.
A significant fraction of the \ttbar background is suppressed by rejecting events with \nsmallr\ $>1$ in the 0L and 1L channels, and \nsmallr\ $>2$ in the 2L channel of the merged-jet
topology analysis, where \nsmallr\ represents the additional \smallr\ jet multiplicity. This requirement is not needed in the 2L channel of the resolved-jet topology analysis where
the top quark background is negligible.

The dominant background that remains after the application of the event selection described above is \PV{}+jets.
The contribution from this background is suppressed by requiring the dijet invariant mass $m_{\text{jj}}$ (calculated using two \smallr\ jets)
in the resolved-jet topology analysis and \msd\ in the merged-jet topology analysis to satisfy $50<m_{\text{jj}}(\msd)<200\GeV$.
Contributions from \ttbar and single top processes remain significant in the 0L and 1L channels
because of the presence of at least one \PW boson and because \bq\ quarks are often
misidentified as \cq\ quarks by the \PQc tagging algorithms.
Contributions from diboson processes are typically small as a result of their small production cross sections.
The background originating from \PH bosons decaying into \PQb quarks presents kinematic properties similar to those of the signal,
with the exception of a higher average energy of neutrinos in \bq jets than in \cq jets.
This background is reduced by exploiting dedicated jet flavour taggers, as described in Sections~\ref{sec:resolvedanalysis} and \ref{sec:boostedanalysis}.

The details of the resolved-jet topology and merged-jet topology analyses are described in Sections~\ref{sec:resolved_an} and~\ref{sec:boosted_an}, respectively.
Section~\ref{sec:systematics} is dedicated to the treatment of the systematic uncertainties and Section~\ref{sec:results} presents the results of the two analyses and of their combination.
Section~\ref{sec:results} presents also the strategy that is used to make the two analyses mutually exclusive in order to facilitate their combination for the final results.

\section{Resolved-jet topology analysis \label{sec:resolvedanalysis}}
\label{sec:resolved_an}

Approximately 95\% of the $\PV\PH$ events produced at $\sqrt{s}=13\TeV$ have a vector boson with \pt lower than 200\GeV, corresponding to the phase space region where
the \PH boson decay products generally give rise to two distinctly reconstructed \smallr\ jets in the CMS detector.
The resolved-jet analysis aims to exploit a large fraction of this phase space which, however, contains a sizeable background contamination.
The requirement of a moderate boost of the vector boson is then found to be crucial to the reduction of \PV{}+jets and \ttbar backgrounds. Dedicated charm taggers based on
ML are used to order the jets in the event by their likelihood to be \cq jets that are considered for use in reconstructing the \PH boson candidate.

Backgrounds arise from the production of \PW and \PZ bosons in association with one or more jets, single and pair-produced top quarks, and
diboson events. A small residual QCD background is present in the 0L and 1L channels. High-purity control regions for the \Vudsg
and \ttbar backgrounds are identified in data and used to estimate expected yields of these backgrounds in the signal region. Samples of events in regions that are disjoint
from the signal region in \cq tagging probability and dijet mass but which are enhanced in \Vbb\ and
\Vcc\ production are used to provide data-driven constraints on the \Vbb\ and \Vcc\ backgrounds, respectively. Finally, a binned maximum
likelihood fit is carried out simultaneously in the signal region and in the control regions for all channels.

\subsection{Higgs boson reconstruction}
\label{sec:HrecoResolved}

\begin{figure}
  \centering
  \includegraphics[width=0.65\textwidth]{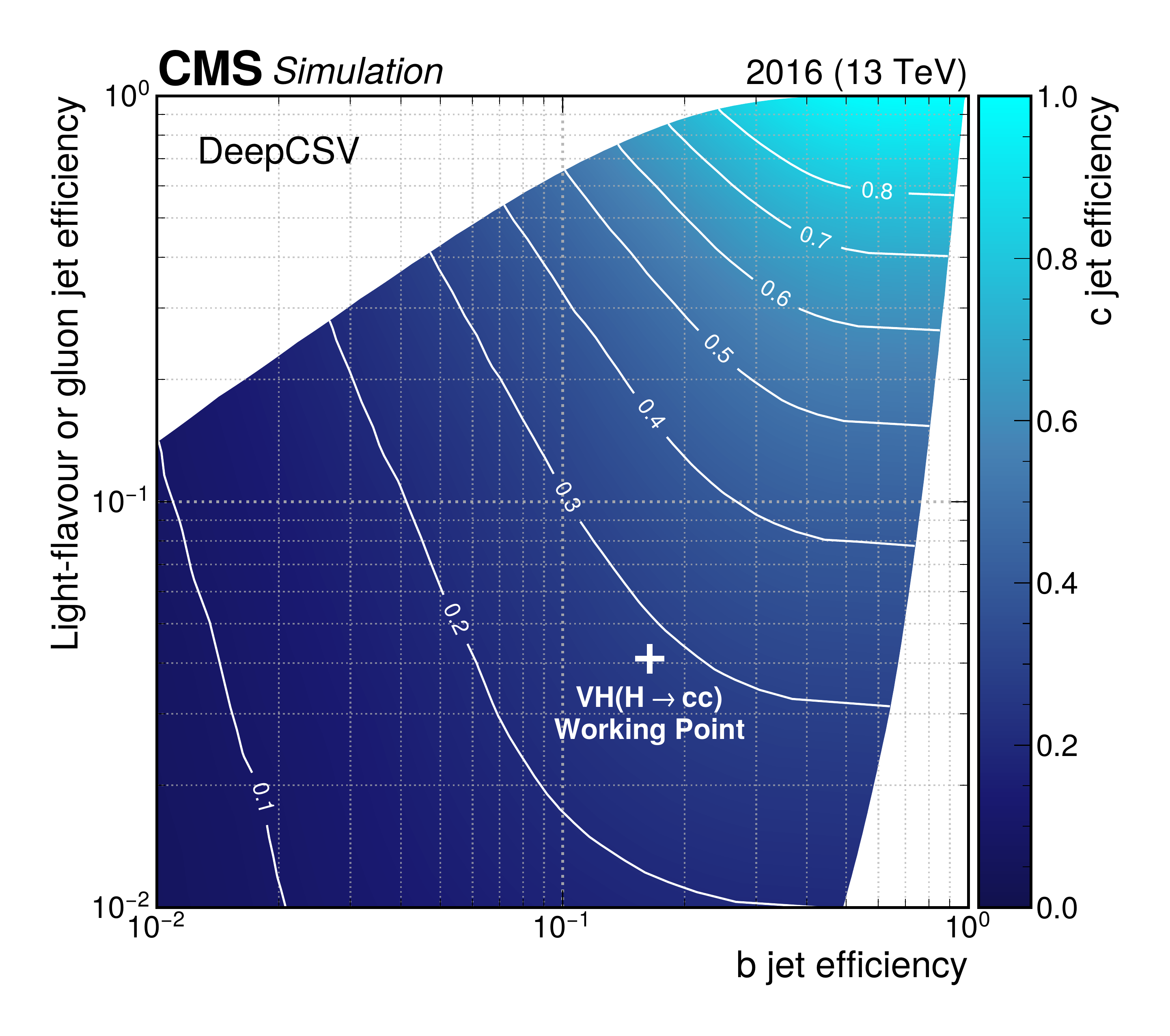}
  \caption{Efficiency to tag a \cq jet as a function of the \bq jet and light-flavour quark or gluon jet mistag rates. The working point adopted in the resolved-jet topology analysis
to select the leading \textit{CvsL} jets is shown with a white cross. The white lines correspond to \cq jet iso-efficiency curves. The plot makes use of jets with $\pt>20$\GeV that have been clustered with AK4 algorithm in a simulated \ttbar{}+jets sample before application of data-to-simulation reshaping scale factors.}
  \label{fig:CTagEfficiency}
\end{figure}

The \PH candidate is reconstructed as two distinct \smallr\ jets. The identification of \cq jets among those arising from other flavours of quarks or gluons is achieved with the Deep Combined Secondary Vertex (DeepCSV) algorithm~\cite{Sirunyan:2017ezt}.
This algorithm encodes a multiclassifier based on advanced ML techniques and provides three output weights $p(\cPqb)$, $p(\PQc)$, and $p(\cPqu{}\cPqd{}\cPqs{}\Pg)$ which can be interpreted as
the probabilities for a given jet to have originated from a bottom quark, a charm quark, or a gluon or light-flavour quark, respectively.
By combining the various DeepCSV outputs, it is possible to define two discriminators for \cq tagging. The inputs to the DeepCSV algorithm are variables constructed from observables associated with the reconstructed primary and secondary vertices, tracks, and jets.
The discrimination between \cq jets and light-flavour quark or gluon jets is achieved via the probability ratio defined as $\textit{CvsL} = p(\PQc) / [ p(\PQc) + p(\cPqu{}\cPqd{}\cPqs{}\Pg) ]$.
In the same way, discrimination between \cq jets and \bq jets makes use of the
probability ratio defined as $\textit{CvsB} = p(\PQc) / [ p(\PQc) + p(\cPqb) ]$. The two discriminator
ratio values for each jet define a two-dimensional distribution. The
resulting \cq tagging efficiency as a function of the \bq jet and light-flavour
quark or gluon jet efficiencies is shown in Fig. \ref{fig:CTagEfficiency}.
To account for residual O(10\%) differences in the distributions of $\textit{CvsL}$ and $\textit{CvsB}$ found in the comparison of data and simulation, reshaping scale factors have been extracted using an iterative fit to  distributions in control regions enriched in Drell--Yan+jets, semileptonic \ttbar{}+jets, and \PW{}+\PQc events that provide data samples with large fractions of light-flavour quark or gluon jets, \bq jets, and \cq jets, respectively.
The corresponding uncertainties, evaluated on a per jet basis as a function of the jet flavour, range from 2\% for bottom, gluon, and light-flavoured quark jets to 5\% for \PQc jets.

The probability ratios $\textit{CvsL}$ and $\textit{CvsB}$ are used to discriminate candidates that are
consistent with the \cq jet hypothesis from jets originating from light-flavour quarks or gluons,
and \bq quarks, respectively. The two jets with the highest score of $\textit{CvsL}$ in the event are chosen
to build the \PH candidate four-vector. Events are required to have the leading jet in $\textit{CvsL}$ score passing
the \cq tagger working point requirements $(\textit{CvsL}>0.4,~\textit{CvsB}>0.2)$. This working point has been chosen such that the efficiency to identify a \cq jet is
approximately 28\%, while the misidentification rate is 4\% for light-flavour quark or gluon jets and 15\% for \bq jets.
The misidentification rate of $\tau$ leptons as \cq jets is larger than the misidentification rate of \bq jets as \cq jets. However, the kinematic properties of the $\tau$ decays are exploited by the BDTs, as described
in Section~\ref{sec:ResoStrategy}, to discriminate signal \cq jet pairs from misidentified jet pairs coming from $\tau$ leptons. The contribution of the \vhtautau process in the high signal purity bins of the BDT
distributions has a negligible impact on the final results. To account for jets originating from final-state radiation (FSR), additional jets with
$\pt > 30\GeV$ and $\abs{\eta}<3.0$ are included in the calculation of the components of the \PH candidate four-vector if they lie
in a cone of $\Delta R<0.8$ centred on the direction of one of the two leading jets.

\subsection{Signal extraction}
\label{sec:ResoStrategy}
In addition to the selections reported in Sections \ref{sec:selection} and \ref{sec:HrecoResolved}, in the 1L and 0L channels of the resolved-jet topology analysis, where larger backgrounds are expected,
the \PV\ candidates are required to have \pt of at least 100 and 170\GeV respectively, while, for the same channels, the \PH candidates are required to have \pt
of at least 100 and 120\GeV. In the 2L channel, where the background from \ttbar production is much smaller and the effective signal cross section is also lower,
two regions are considered: a low-\ptV region defined by $50 < \ptV < 150\GeV$ and a high-\ptV region with $\ptV > 150$\GeV (no upper cut is applied on $\ptV$).
In \vhcc\ signal events, the vector boson is typically produced in the azimuthal direction opposite to that of the \PH boson.
Therefore, an additional requirement on the difference in azimuthal angle between the reconstructed \PV
and \PH candidate, \dphiVH\ $>$2.5 ($>$2.0 in the 0L channel), is applied.

In the signal regions defined by the application of the selection criteria mentioned above,
a boosted decision tree (BDT) with gradient boost~\cite{Hocker:2007ht} has been trained to enhance the signal separation from background. The same simulated samples,
each normalised to the cross section of the relevant physics process, have been split into two independent subsets used for training and testing the BDTs.
All the data-to-simulation scale factors relative to trigger efficiency, lepton identification and isolaton efficiency, and c-tagging
have been applied to the simulated samples. Separate BDTs have been trained for 0L, 1L, and 2L (low-\ptV and high-\ptV) channels.
The muon and electron samples were combined to train the BDTs to benefit from a higher number of simulated events.
The distributions of all variables used to construct the BDT discriminator and hence the BDT distribution itself are taken from simulation after the application of the corrections detailed in Section~\ref{sec:mcsamples}.
Table~\ref{tab:ResoBDTInputs} lists the input variables considered in each channel. As expected, the most discriminating variables are found to be the \PH candidate invariant mass and the $\textit{CvsL}_{\text{max}}$.

\begin{table*}[htbp]
\topcaption{Variables employed in the training of the BDT used for each channel of the resolved-jet topology analysis.
The 2L case has separate training for the low- and high-\ptV channels, but exploits the same input variables.}
\label{tab:ResoBDTInputs}
\centering
\resizebox*{\textwidth}{!}{
\begin{tabular}{llccc}
\hline
Variable & Description & 0L & 1L & 2L \\
\hline
$m(\PH)$ & \PH mass                                                   & \xmark & \xmark & \xmark \\
\pt(\PH) & \PH transverse momentum                                    & \xmark & \xmark & \xmark \\
\ptV & vector boson transverse momentum                                     & \xmark & \xmark & \xmark \\
$m(\PV)$ & vector boson mass                                                 & \NA    & \NA    & \xmark \\
$\mT(\PV)$ & vector boson transverse mass                                    & \NA    & \xmark & \NA    \\
\ptmiss & missing transverse momentum                                         & \xmark  & \xmark & \NA   \\
$\ptV/\pt(\PH)$ & ratio between vector boson and \PH transverse momenta    & \xmark & \xmark & \xmark \\
$\textit{CvsL}_{\text{max}}$ & $\textit{CvsL}$ value of the leading $\textit{CvsL}$ jet    & \xmark & \xmark & \xmark \\
$\textit{CvsB}_{\text{max}}$ & $\textit{CvsB}$ value of the leading $\textit{CvsL}$ jet    & \xmark & \xmark & \xmark \\
$\textit{CvsL}_{\text{min}}$ & $\textit{CvsL}$ value of the subleading $\textit{CvsL}$ jet    & \xmark & \xmark & \xmark \\
$\textit{CvsB}_{\text{min}}$ & $\textit{CvsB}$ value of the subleading $\textit{CvsL}$ jet    & \xmark & \xmark & \xmark \\
$\pt{}_{\text{max}}$ & \pt of the leading $\textit{CvsL}$ jet    & \xmark & \xmark & \xmark \\
$\pt{}_{\text{min}}$ & \pt of the subleading $\textit{CvsL}$ jet    & \xmark & \xmark & \xmark \\
\dphiVH & azimuthal angle between vector boson and \PH                  & \xmark & \xmark & \xmark \\
$\Delta R(\jone,\jtwo)$ & $\Delta R$ between leading and subleading $\textit{CvsL}$ jets & \NA  & \xmark & \xmark \\
$\Delta\phi(\jone,\jtwo)$ & azimuthal angle between leading and subleading $\textit{CvsL}$ jets & \xmark & \xmark & \NA    \\
$\Delta\eta(\jone,\jtwo)$ & difference in pseudorapidity between leading and subleading $\textit{CvsL}$ jets & \xmark & \xmark & \xmark \\
$\Delta\phi(\ell_1,\ell_2)$ & azimuthal angle between leading and subleading \pt leptons & \NA   & \NA   & \xmark \\
$\Delta\eta(\ell_1,\ell_2)$ & difference in pseudorapidity between leading and subleading \pt leptons &  \NA  &  \NA  & \xmark \\
$\Delta\phi(\ell_{1},\jone)$ & azimuthal angle between leading \pt lepton and leading $\textit{CvsL}$ jet & \NA  & \xmark & \NA  \\
$\Delta\phi(\ell_{2},\jone)$ & azimuthal angle between subleading \pt lepton and leading $\textit{CvsL}$ jet & \NA  & \NA  & \xmark \\
$\Delta\phi(\ell_2,\jtwo)$ & azimuthal angle between subleading \pt lepton and subleading $\textit{CvsL}$ jet & \NA  &  \NA  & \xmark \\
$\Delta\phi(\ell_1,\met)$ & azimuthal angle between leading \pt lepton and missing transverse momentum &  \NA  & \xmark & \NA  \\
\nsmallr\ & number of \smallr\ jets minus the number of FSR jets                              & \xmark & \xmark & \xmark \\
$N_{5}^{soft}$ & multiplicity of soft track-based jets with $\pt>5$\GeV                  & \xmark & \xmark & \xmark \\
\hline
\end{tabular}
}
\end{table*}

The remaining background contribution is estimated from a combination of simulated events and data.
While the normalisations of QCD, single-top, diboson, and \vhbb
processes are estimated via simulation, the normalisations of the \PV{}+jets and \ttbar{}+jets backgrounds
are determined from fits to data in dedicated control regions in order to avoid potential mismodelling
of the flavour composition of these samples.
Four control regions per channel are designed to constrain the most important background processes: a region dominated
by \ttbar{}+jets events (TT), a region targeting \PV{}+jets with at least one jet originating from light-flavour quarks or gluons (LF),
a region enriched in \PV{}+jets events with one \PQb jet and one \PQb or \PQc jet (HF), and a region
enriched with \Vcc\ events (CC). The definitions of the different control regions are based mainly on the inversion of the criteria
on the charm tagger discriminators values of the $\textit{CvsL}$-leading jet applied to define the signal regions.
To define the LF control region the selection $(\textit{CvsL}<0.4,~\textit{CvsB}>0.2)$ is used while both the HF and TT control regions
are defined applying the selection $(\textit{CvsL}>0.4,~\textit{CvsB}<0.2)$. In order to differentiate the TT from HF control regions, further requirements
are applied such as a veto on the reconstructed \PZ boson mass in the 2L channel, $m_{\ell\ell}\notin [75,120]\GeV$, and the requirement $\nsmallr\geq2$.
The CC control region is defined identically to the signal region, except for inverting the requirement on the \PH mass.
The simulated \PV{}+jets backgrounds are similarly split into four classes depending on the flavour(s) of
the additional jet(s) present in the processes: \PV{}+2 light-flavour quark or gluon jets, \Vudsg and 1\PQb or 1\PQc, \PV{}+$\PQb\PQb$ or $\PQb\PQc$, and \PV{}+$\PQc\PQc$ jets.

Separate normalisation scale factors are used to constrain \PZ{}+jets processes in the 0L and 2L channels, while the
normalisation scale factors related to \PW{}+jets processes are shared between the 0L and 1L analysis channels.
To constrain the \ttbar{}+jets process, on the other hand, each channel relies on its own independent normalisation scale factors.
The normalisation scale factors are measured together with the signal strength modifier through a simultaneous fit to data in all
control and signal regions for all of the analysis channels. The simulated diboson background is split according to the presence or absence of a \PZ boson
decaying to a pair of charm quarks, labelling them as \vzcc if such a \PZ boson decay is present and VV+other otherwise.
Whereas in the signal regions the BDT discriminator is used for the final signal extraction, in the control regions the shape of
the $\textit{CvsB}$ distribution is used in the TT, HF, and CC regions, while that of $\textit{CvsL}$ is used in the LF region.
The reason of this choice lies in the fact that $\textit{CvsB}$ provides the best discriminant between \PQb and \PQc jets and thus it is used in the control regions where there is
 an enhanced presence of \PQb jets, while $\textit{CvsL}$ is more efficient in separating light-flavour quark or gluon jets from \PQc jets and hence it is preferred in the LF control region.

\begin{figure}[!hp]
  \centering
  \includegraphics[width=0.375\textwidth]{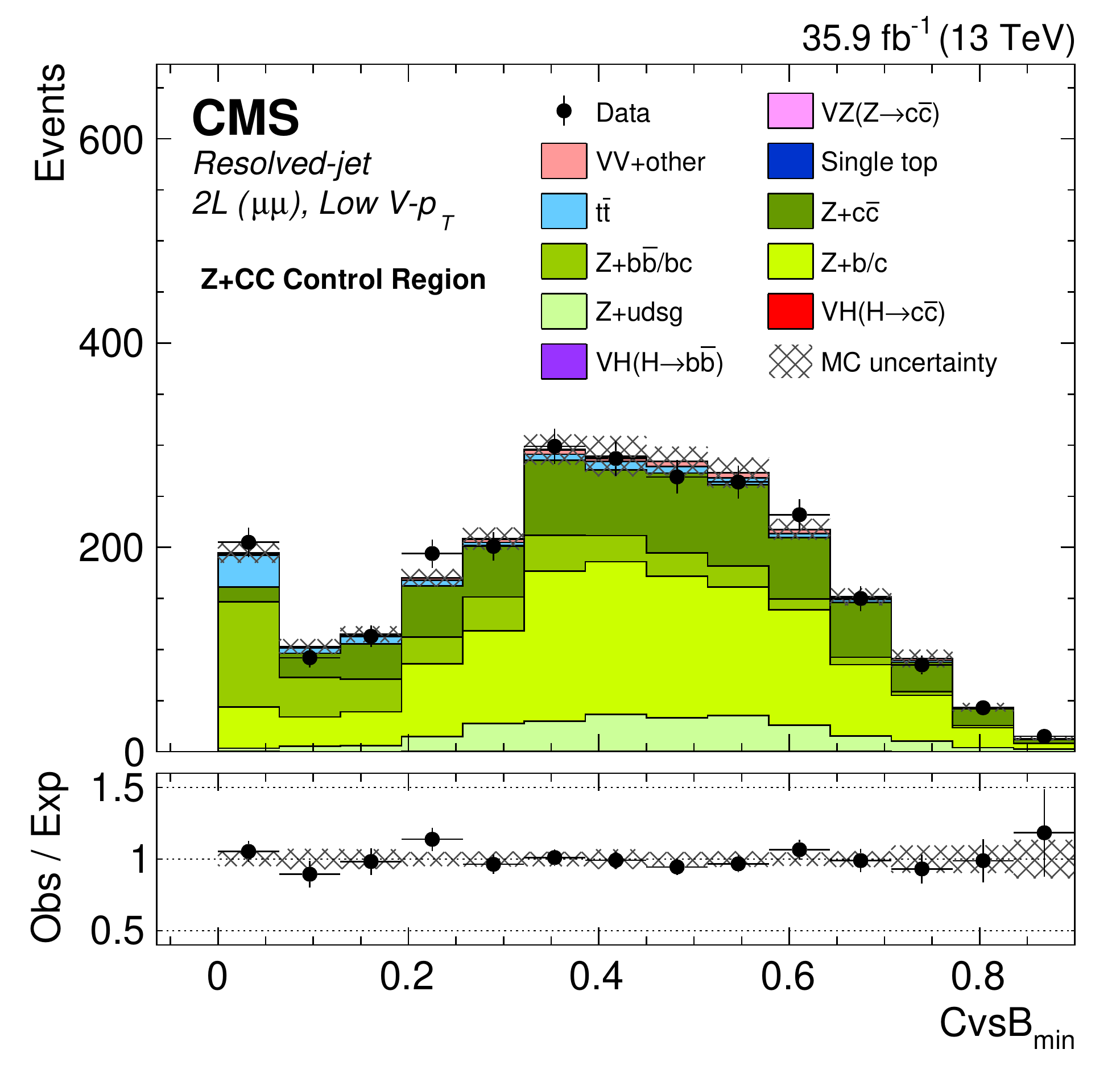}
  \includegraphics[width=0.375\textwidth]{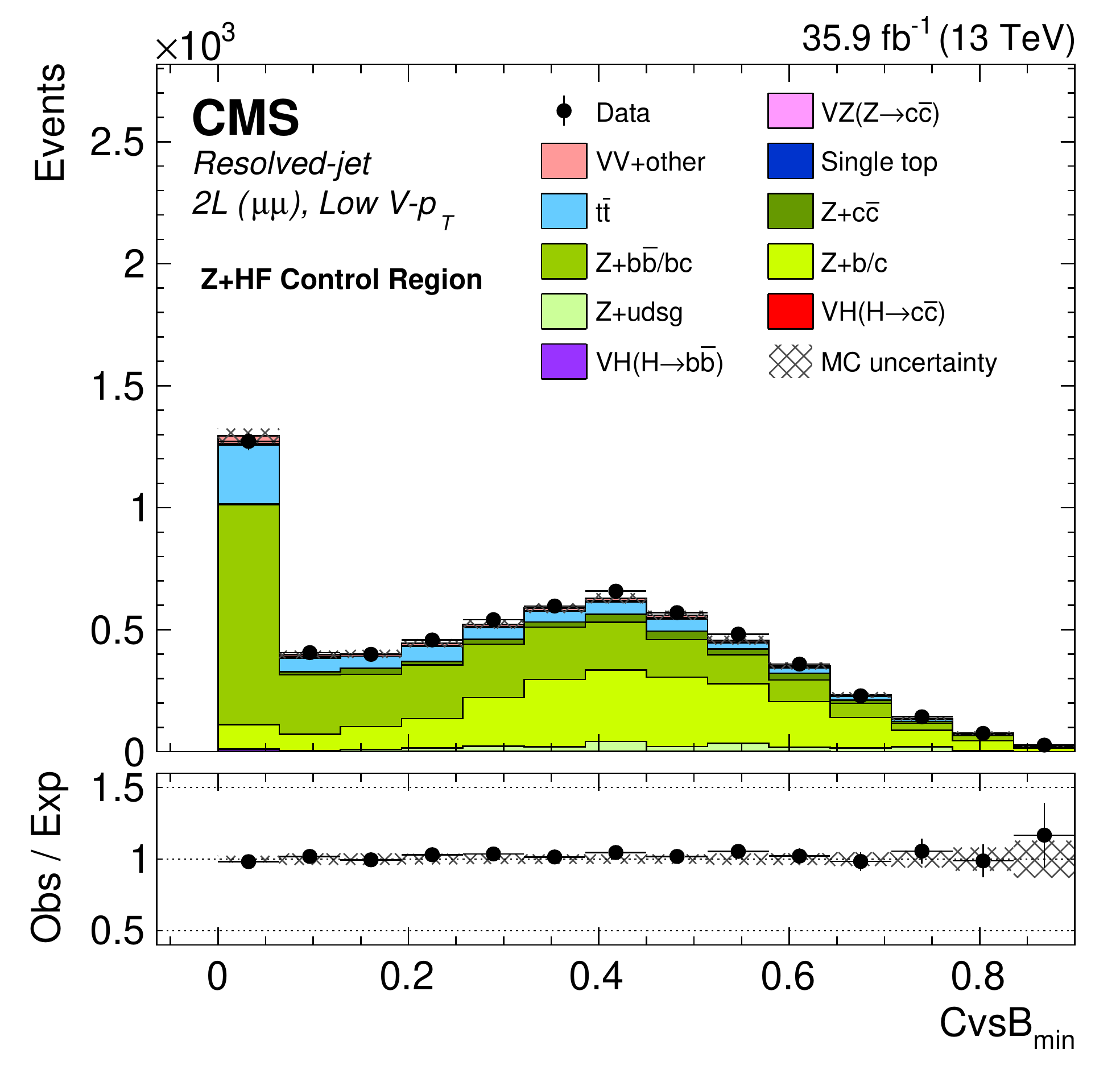}
  \includegraphics[width=0.375\textwidth]{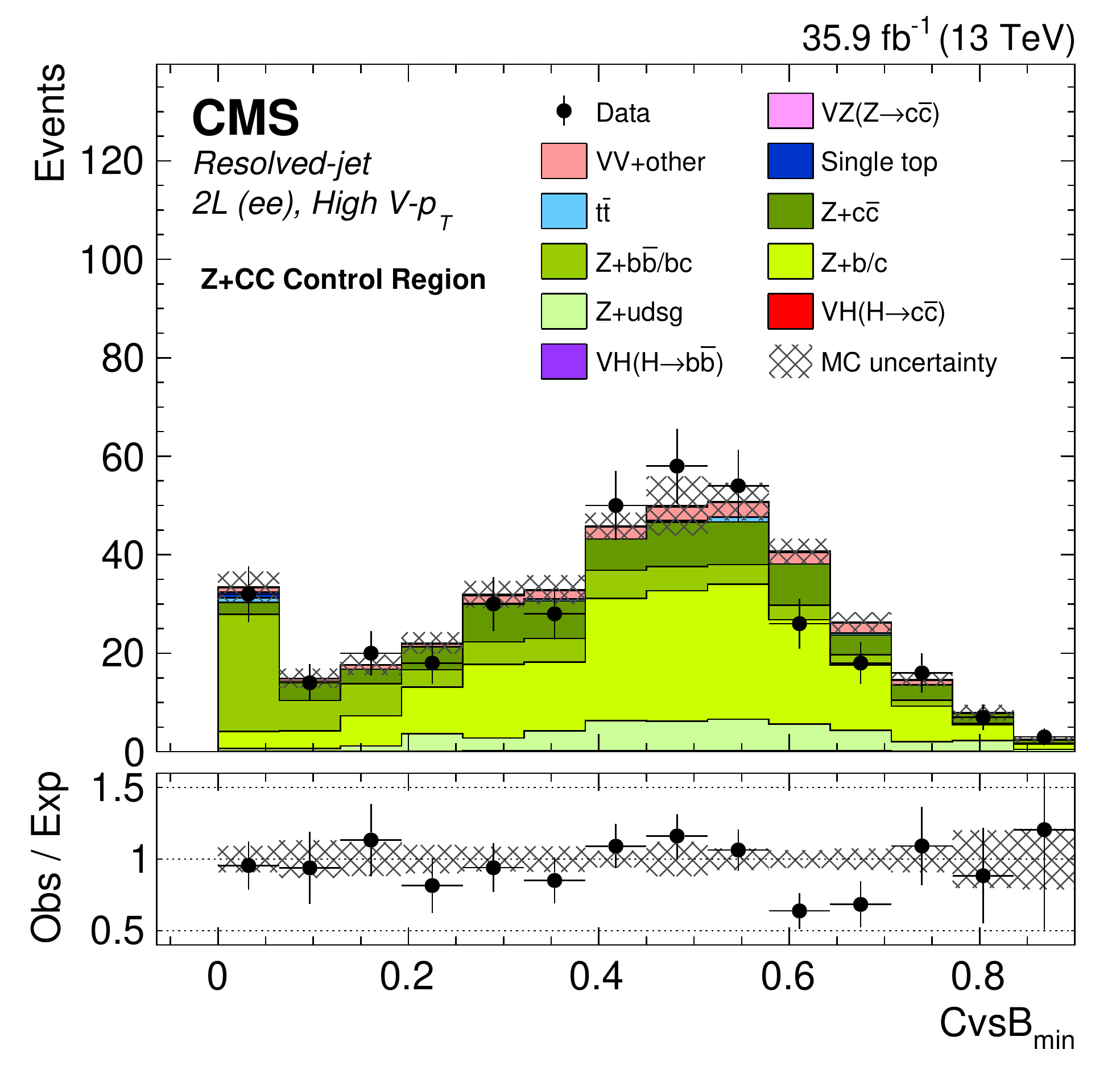}
  \includegraphics[width=0.375\textwidth]{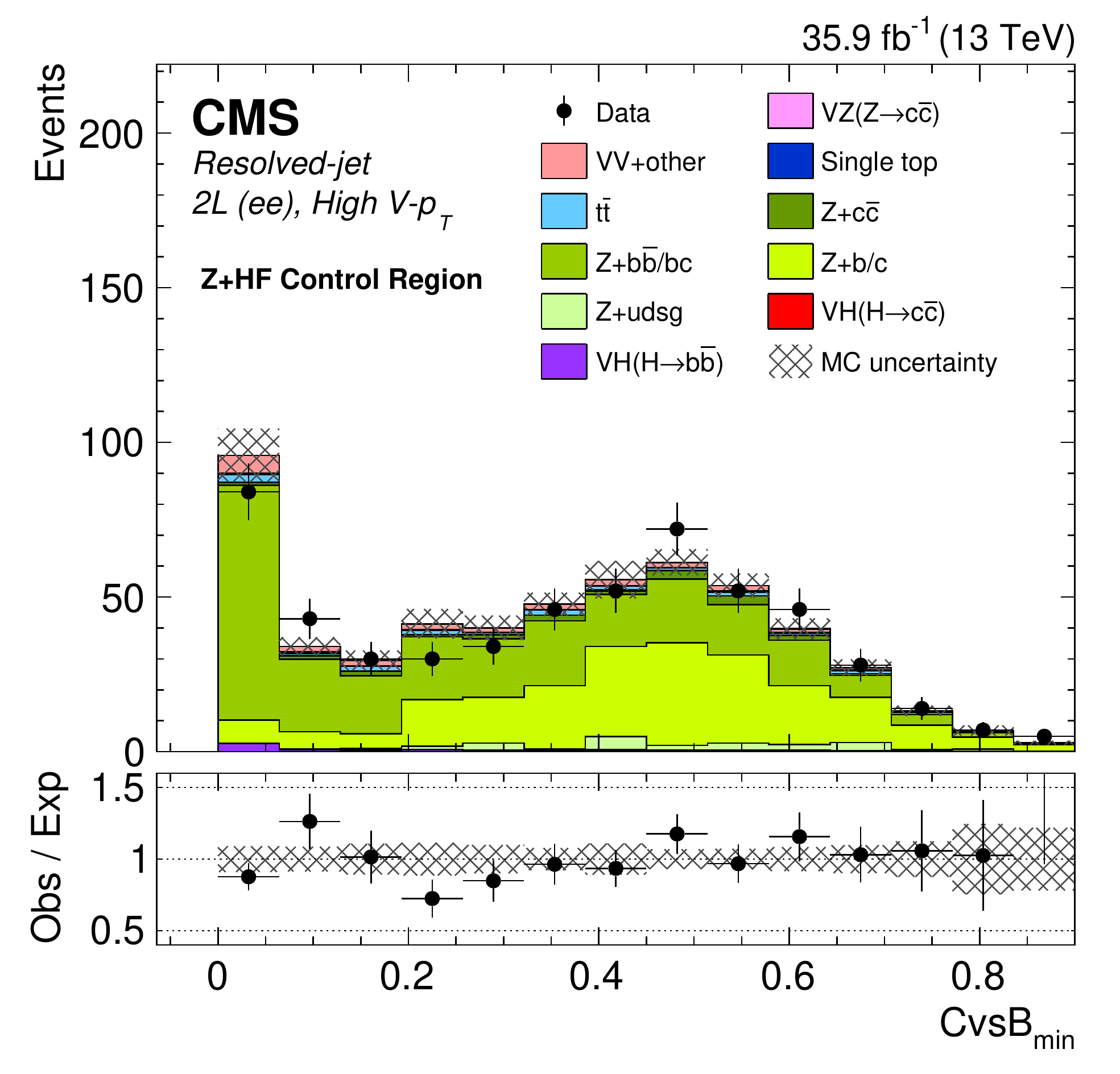}
  \includegraphics[width=0.375\textwidth]{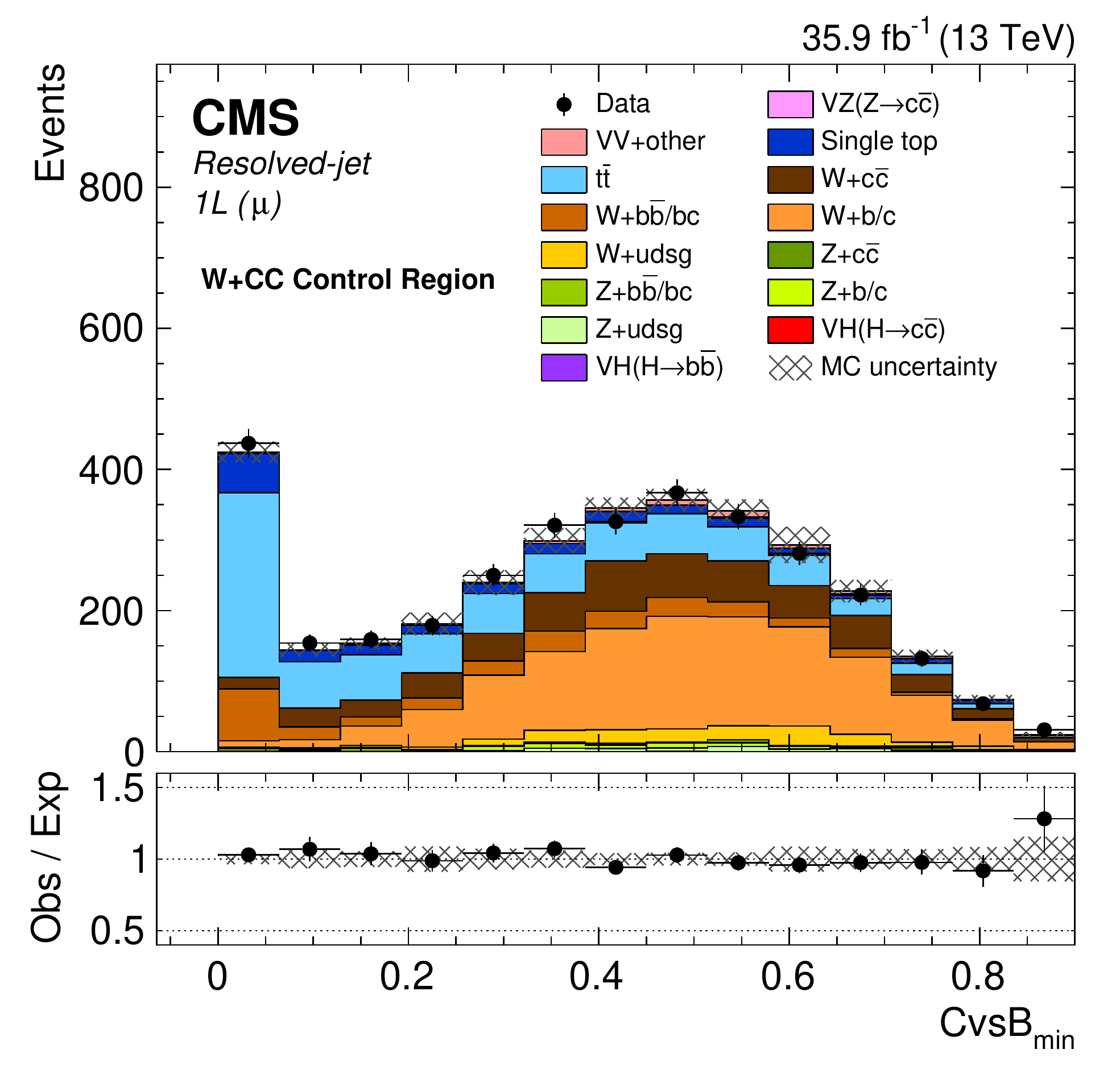}
  \includegraphics[width=0.375\textwidth]{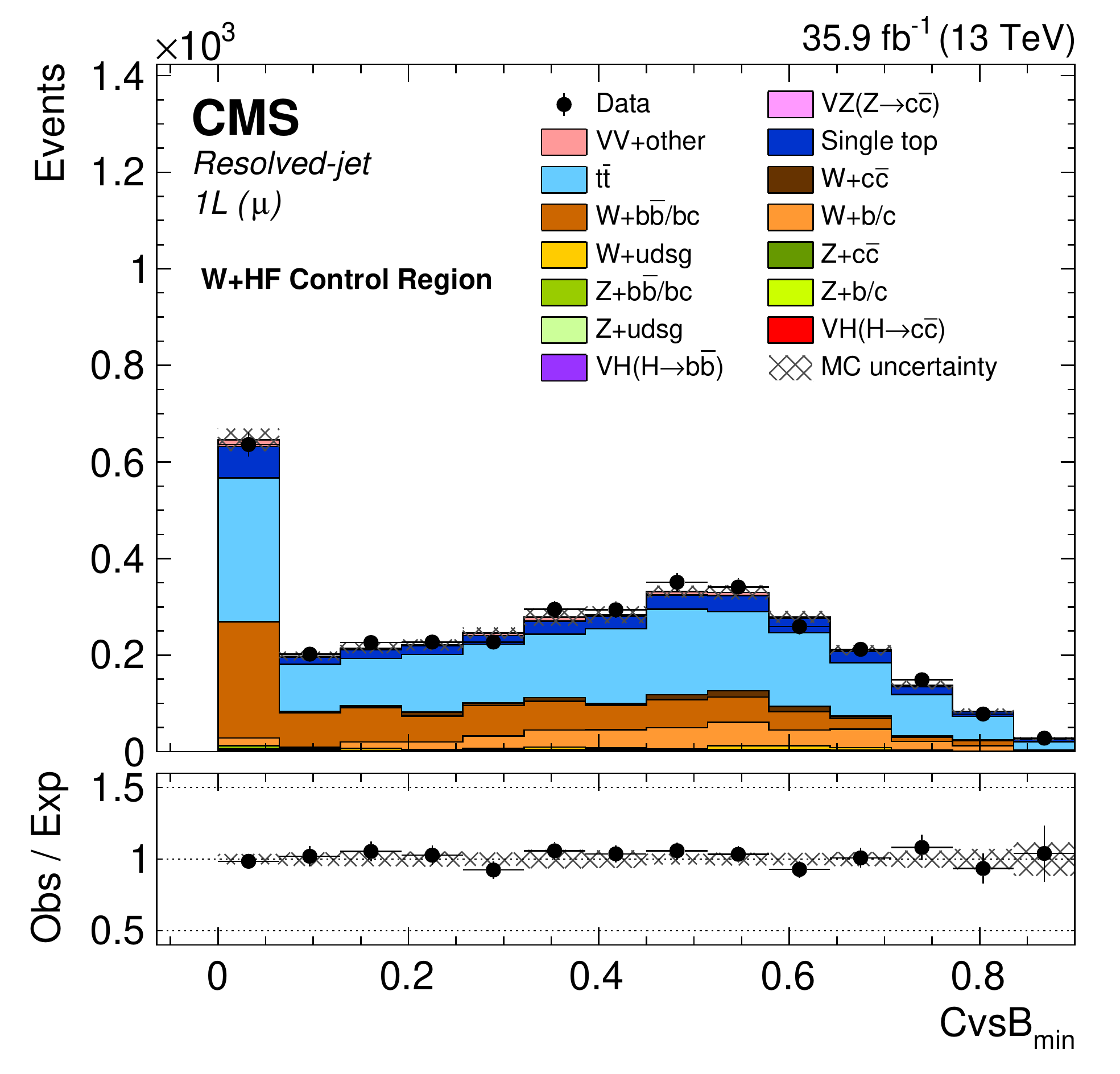}
  \includegraphics[width=0.375\textwidth]{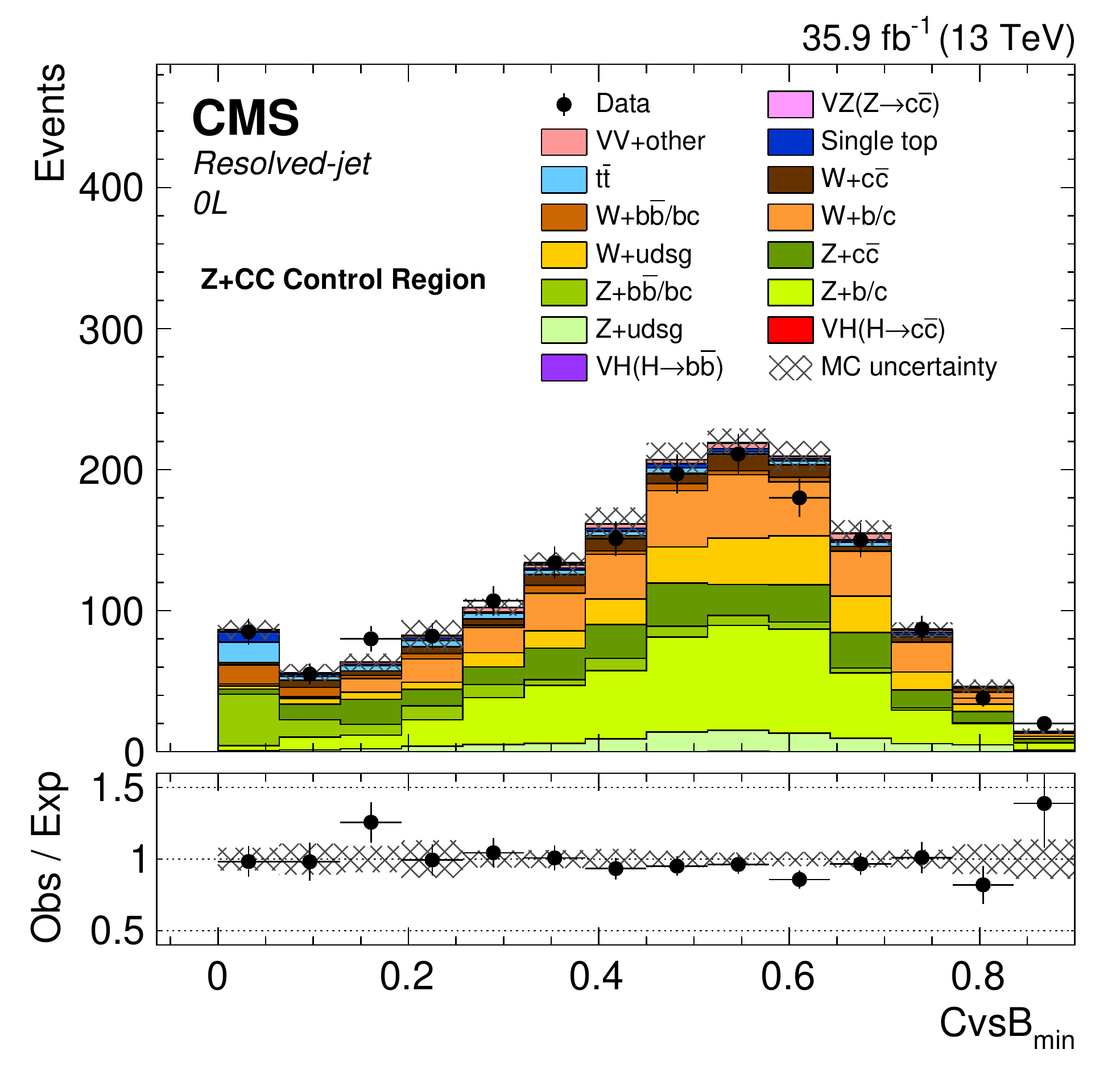}
  \includegraphics[width=0.375\textwidth]{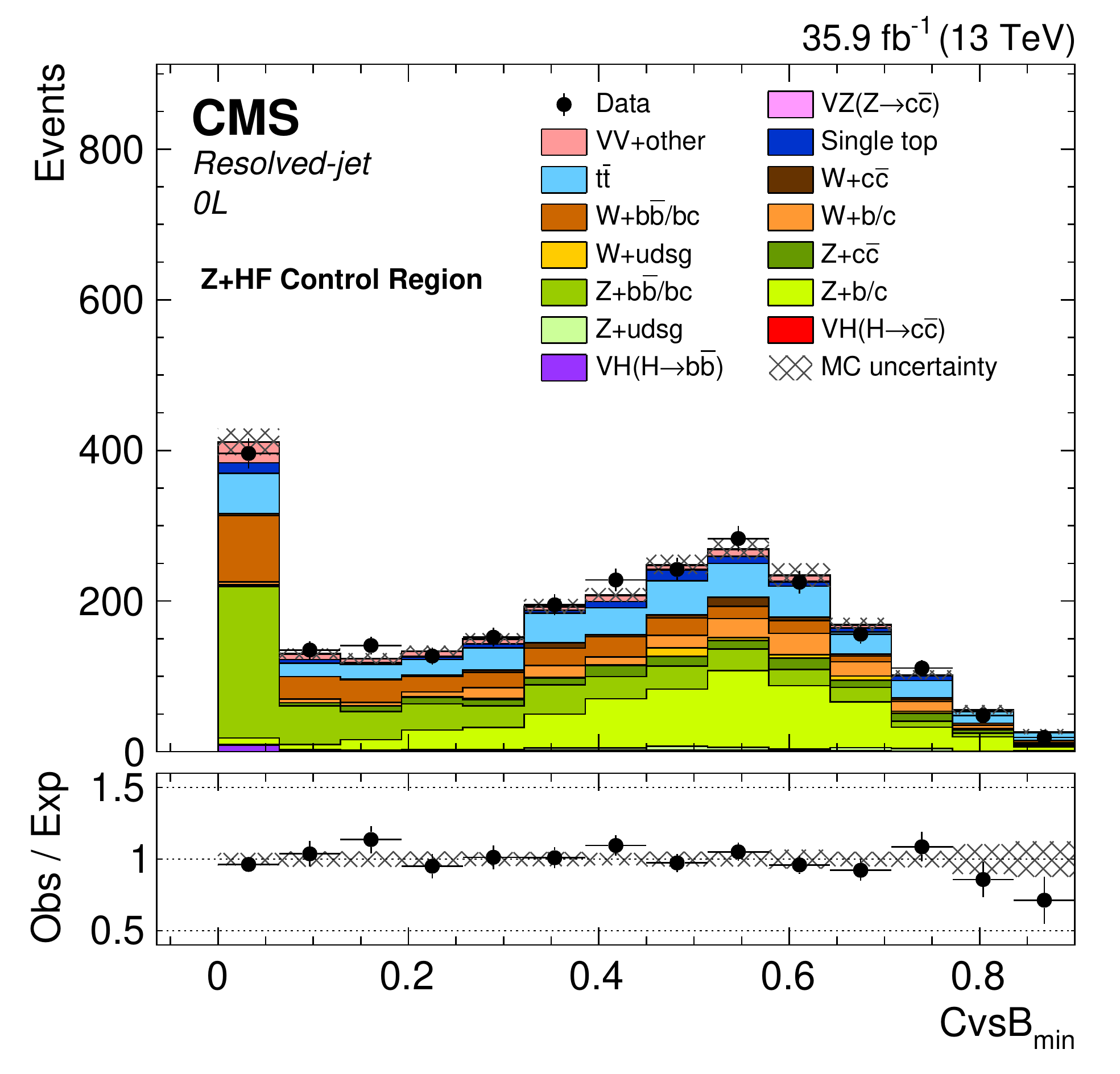}
  \caption{Post-fit $\textit{CvsB}_{\text{min}}$ distributions in the CC (left panel) and HF (right panel) control regions for the 2L (\Zmm) low-\ptV, 2L (\Zee) high-\ptV, 1L (\Wmn), and 0L channels.}
  \label{fig:BDTCR2lepLow}
\end{figure}

Figure~\ref{fig:BDTCR2lepLow} shows the distributions of the $CvsB$ discriminant for the subleading $\textit{CvsL}$ jet for the HF and CC control regions in the 2L (\Zmm) low-\ptV, 2L (\Zee) high-\ptV, 1L (\Wmn), and 0L channels.
The post-fit distributions (for fit details, see Section~\ref{sec:results}) in Fig.~\ref{fig:BDTCR2lepLow} show good agreement between the data and the simulation in these two most
significant control regions. Moreover, the employment of the full distribution of the $\textit{CvsB}$ score provides
a good separation between the \PV{}+$\PQb\PQb$ and \PV{}+$\PQc\PQc$ processes that makes it possible to constrain these two backgrounds.
The corresponding distributions for the other channels are not shown but are similar in their behaviour.

\section{Merged-jet topology analysis \label{sec:boostedanalysis}}
\label{sec:boosted_an}
For the case of a Lorentz-boosted \PH boson as flagged by a \PV boson with $\ptV \gtrsim 200\GeV$, a merged-jet topology is considered.
The dominant backgrounds after the baseline selection presented in Section~\ref{sec:selection} come from \PV{}+jets and \ttbar processes.
The \PV bosons in the signal process have on average larger \pt than those from the \PV{}+jets background. The analysis focuses on the reconstruction of moderately to highly Lorentz-boosted \PH bosons where the decay products are contained in a single \larger\ jet. Dedicated object reconstruction tools based on \larger\ jets and advanced ML techniques were developed to identify and reconstruct Lorentz-boosted \PH bosons decaying to charm quarks.

\subsection{Higgs boson reconstruction}
The cornerstone of the merged-jet topology analysis is the reconstruction of the \HCC\ candidate in a single, \larger\ jet, which has the potential to provide a better signal purity because the signal has a tendency to be more boosted than the dominant \PV{}+jets and \ttbar backgrounds, as noted above. In view of this, the high-\pt regime with $\ptV \gtrsim 200\GeV$, though representing no more than approximately 5\% of the total phase space, can provide a significant contribution to the search. Moreover, the merged-jet approach has important advantages over the resolved-jet approach at high \pt.  The possibility for both \cq\ quarks to reside in a single \larger\ jet enhances the signal acceptance, improves the identification of the correct pair of jets to use in reconstructing the \PH boson, and similarly facilitates the task of taking into account any FSR that may have been emitted by the quarks.  A more detailed discussion of the potential advantages of this approach can be found in Refs.~\cite{Butterworth:2008iy,Plehn:2009rk}.
Given the small fraction of signal events that survive a selection with $\ptV \gtrsim 200\GeV$, it is critical to carefully choose the $R$ parameter of the jet clustering algorithm. In general, the angular separation between the decay products of a Lorentz-boosted particle such as the \PH boson is approximately given by $\Delta R \sim 2m_{\PH}/\ptH$. For a $\ptH\sim\ptV$ of 200\GeV, this gives $\Delta R \approx 1.25$. For good signal purity and acceptance, we have thus chosen to use \larger\ jets clustered by the anti-\kt algorithm with a distance parameter of $R=1.5$.

As for the resolved-jet topology analysis, one of the biggest challenges is the efficient reconstruction of the pair of \cq\ quarks from the \PH boson decay, while also achieving significant rejection of both light-flavour quarks and gluons, as well as \bq\ quarks that contribute backgrounds to this search.
To this end, a novel algorithm, DeepAK15~\cite{CMS-PAS-JME-18-002}, has been used for the identification of jet substructure to tag \PW, \PZ, and \PH bosons, as well as top quarks.
In addition, DeepAK15 is designed to discriminate between decay modes with different flavour content (e.g. $\HBB$, $\HCC$, $\PH \to \PQq\PAQq\PQq\PAQq$).
The algorithm deploys ML methods on the PF candidates and secondary vertices, which are used as inputs.
DeepAK15 is designed to exploit information related to jet substructure, flavour, and pileup simultaneously,
yielding substantial gains with respect to other approaches~\cite{CMS-PAS-JME-18-002}.
With the use of the adversarial training procedure~\cite{NIPS2017_6699}, the algorithm is largely decorrelated from the jet mass, while preserving most of the method's discriminating power.

\begin{figure}[ht]
\centering
\includegraphics[width=0.45\textwidth]{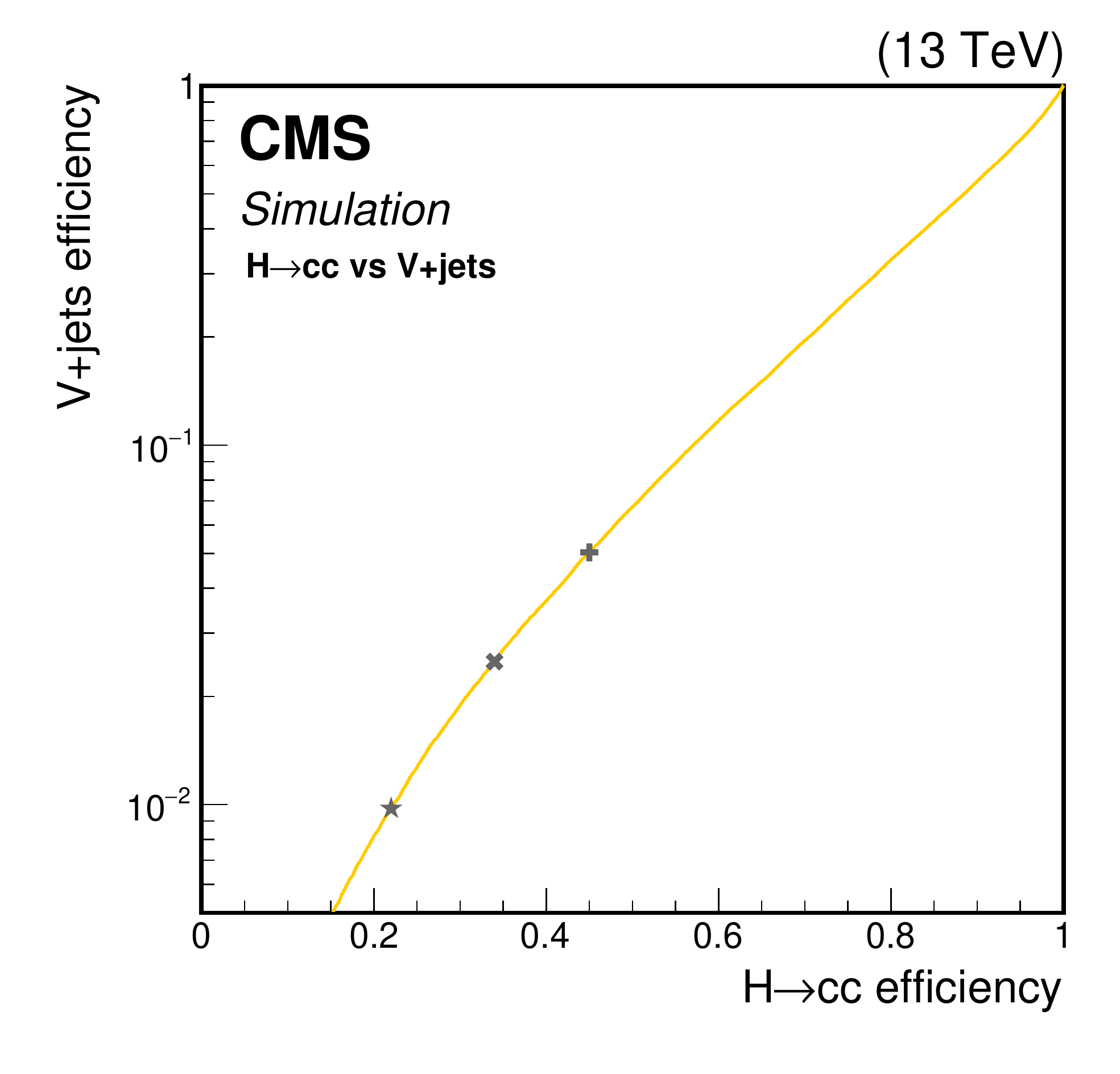}
\includegraphics[width=0.45\textwidth]{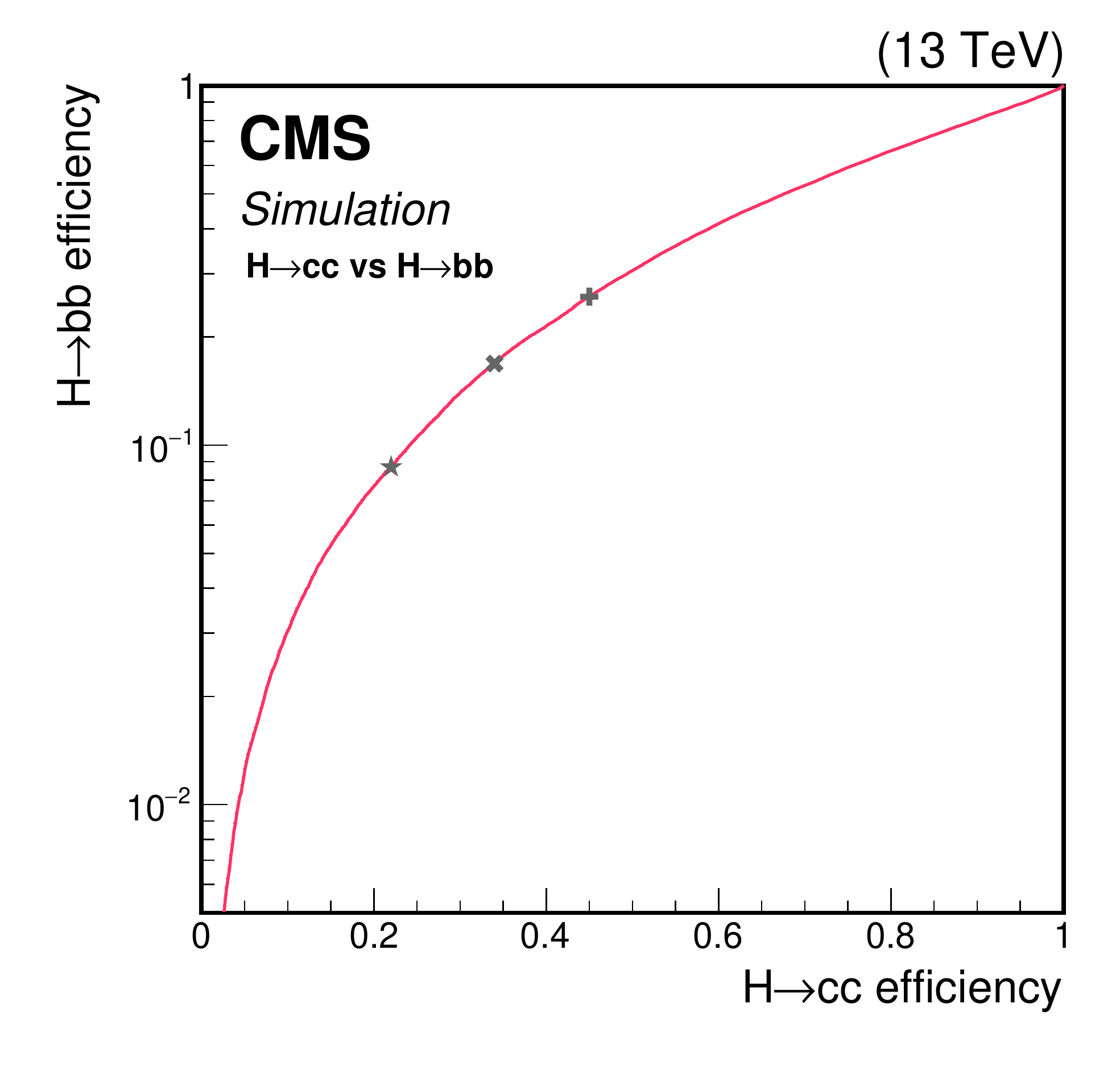}
\caption{\label{fig:roc_deepak15} The performance of the $\ccbar$ discriminant to identify a $\ccbar$ pair in terms of receiver operating characteristic curves, for \larger\ jets with $\pt>200$\GeV, before the application of data-to-simulation scale factors. Left: the efficiency to correctly identify a pair of \cq\ quarks from \PH boson decay vs. the efficiency of misidentifying quarks from the \PV{}+jets process. Right: the efficiency to correctly identify a pair of \cq\ quarks  from \PH boson decay vs. the efficiency of misidentifying a pair of \bq\ quarks from \PH boson decay. The gray stars and crosses on the ROC curves represent the three working points used in the merged-jet topology analysis.}
\end{figure}

The performance in terms of the receiver operating characteristic (ROC) curve of the $\ccbar$ discriminant for identifying a pair of \cq quarks
from \PH boson decay versus quarks from the \PV{}+jets process for \larger\ jets with $\pt>200$\GeV is shown in Fig.~\ref{fig:roc_deepak15} (left).
Three working points (WPs) are defined on the $\ccbar$ tagging discriminant distribution with approximately 1, 2.5, and 5\% misidentification rates,
and the corresponding efficiencies for identifying a $\ccbar$ pair are approximately 23, 35, and 46\%.
Another important parameter is the misidentification of \bq\ jets as signal \cq\ jets.
The corresponding ROC curve is displayed in Fig.~\ref{fig:roc_deepak15} (right).
For the three WPs defined above, the corresponding \bq\ jet misidentification rates are approximately 9, 17, and 27\%.
To improve the sensitivity of the analysis, three mutually exclusive $\ccbar${}-enriched categories, the ``low-purity'' (LP), ``medium-purity'' (MP),
and ``high-purity'' (HP) categories, are defined based on the three WPs. The misidentification rate of $\tau$ leptons as signal \cq jets is larger than the misidentification rate of \bq jets, but typically a factor of two smaller than the  \cq\ jet efficiency. However, due to kinematic and mass selection requirements that will be detailed in Section~\ref{sec:boosted_sigextr}, the \vhtautau contribution is much smaller than the SM \vhcc\ signal and hence has a negligible impact on the final result.
The merged-jet algorithm is calibrated using data and MC simulated samples.
The \pt-dependent data-to-simulation scale factors typically range from 0.85 to 1.30, and the corresponding uncertainties range between 20 and 40\%.

\subsection{Signal extraction}
\label{sec:boosted_sigextr}
In the merged-jet topology analysis, events are required to have at least one \larger\ jet with $\pt>200$\GeV, with the highest \pt\ \larger\ jet selected as the \PH boson candidate.
In \vhcc signal events, the vector boson and the \PH boson are typically emitted back-to-back in $\phi$. Therefore, the difference in azimuthal angle between the reconstructed vector boson and \PH candidate, \dphiVH, is required to be at least 2.5\unit{rad}. To avoid double-counting, \smallr\ jets are removed from the event if they overlap the \PH candidate with $\Delta R(\text{\smallr\ jet}, \PH)<1.5$.

To further distinguish the \VH signal process from the main backgrounds, a separate BDT is developed for each channel. The goal is to define a discriminant that improves the separation between \VH
signal and the main backgrounds, while remaining largely independent of the $\ccbar$ tagging discriminant and the \PH mass. The BDT only makes use of kinematic information from the event,
without including intrinsic properties of \PH such as the flavour content and the mass of the \larger\ jet, which will be used in a fit to the data for the signal extraction.
For the signal process, the \vhbb\ sample was used instead of the \vhcc\ sample, and only events with even event numbers were used to train the BDT while those with odd event numbers were
used for the main analysis. As the BDT is designed to be insensitive to the flavour content of the Higgs candidate, training with the \vhbb\  signal sample results in no loss of performance.
For the background process, only the main backgrounds, e.g., \PZ{}+jets (\PV{}+jets and \ttbar{}+jets) in the case of the 2L (0L and 1L) channel are used.
Table~\ref{tab:kinBDT-inputs} summarises the kinematic variables used as input to the BDT for each of the three channels.

\begin{table*}[htbp]
\topcaption{Variables used in the kinematic BDT training for each channel of the merged-jet topology analysis.}
\label{tab:kinBDT-inputs}
\centering
\resizebox*{\textwidth}{!}{
\begin{tabular}{llccc}
\hline
Variable & Description & 0L & 1L & 2L \\
\hline
\ptV & vector boson transverse momentum                                    & \xmark & \xmark & \xmark \\
\pt(\PH) & \PH transverse momentum                                   & \xmark & \xmark & \xmark \\
$\abs{\eta(\PH)}$ & absolute value of the \PH pseudorapidity             & \xmark & \NA    & \NA    \\
\dphiVH & azimuthal angle between vector boson and \PH                  & \xmark & \xmark & \xmark \\
\met & missing transverse momentum                                         & \NA    & \xmark & \NA    \\
$\Delta\eta(\PH, \ell)$ & difference in pseudorapidity between \PH and the lepton                       & \NA  & \xmark & \NA    \\
$\Delta\eta(\PV,\PH)$ & difference in pseudorapidity between the vector boson and \PH                       & \NA  & \NA    & \xmark \\
$\Delta\eta(\PH, \text{j})$ & min. difference in pseudorapidity between \PH and \smallr\ jets           & \xmark  & \xmark & \xmark \\
$\Delta\eta(\ell, \text{j})$ & min. difference in pseudorapidity between the lepton and \smallr\ jets         & \NA     & \xmark & \NA    \\
$\Delta\eta(\PV, \text{j})$ & min. difference in pseudorapidity between vector boson and \smallr\ jets         & \NA     & \NA    & \xmark \\
$\Delta\phi(\metvec,\text{j})$ & azimuthal angle between \metvec and closest \smallr\ jet                     & \xmark  & \NA    & \NA    \\
$\Delta\phi(\metvec,\ell)$ & azimuthal angle between \metvec\ and lepton                                      & \NA     & \xmark & \NA    \\
$\mT$ & transverse mass of lepton \ptvec + \metvec              & \NA   & \xmark &  \NA  \\
\nsmallr\ & number of additional \smallr\ jets                                             & \xmark & \xmark & \xmark \\
\hline
\end{tabular}
}
\end{table*}

\begin{figure}[!hp]
\centering
\includegraphics[width=0.4\textwidth]{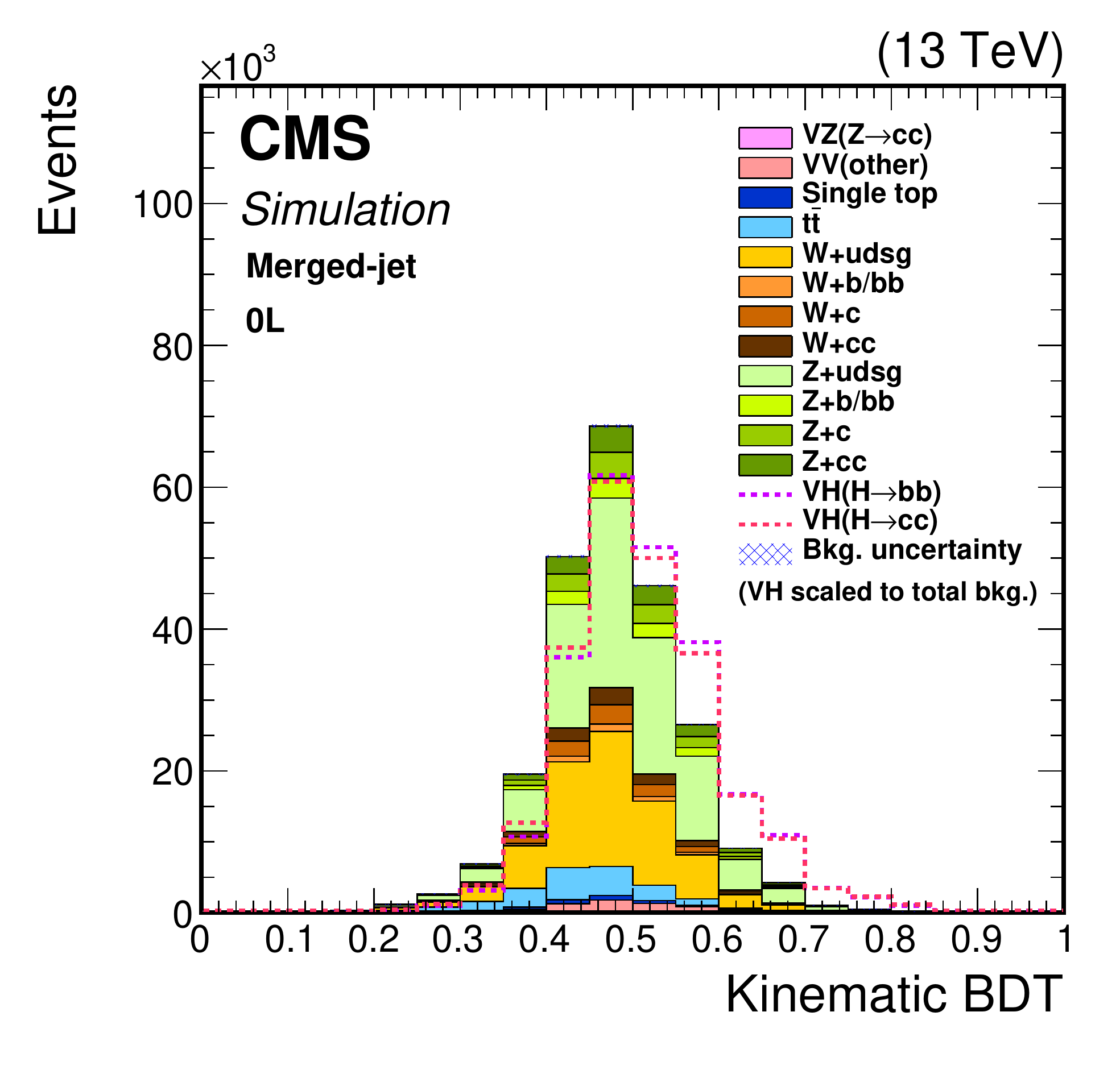}
\includegraphics[width=0.4\textwidth]{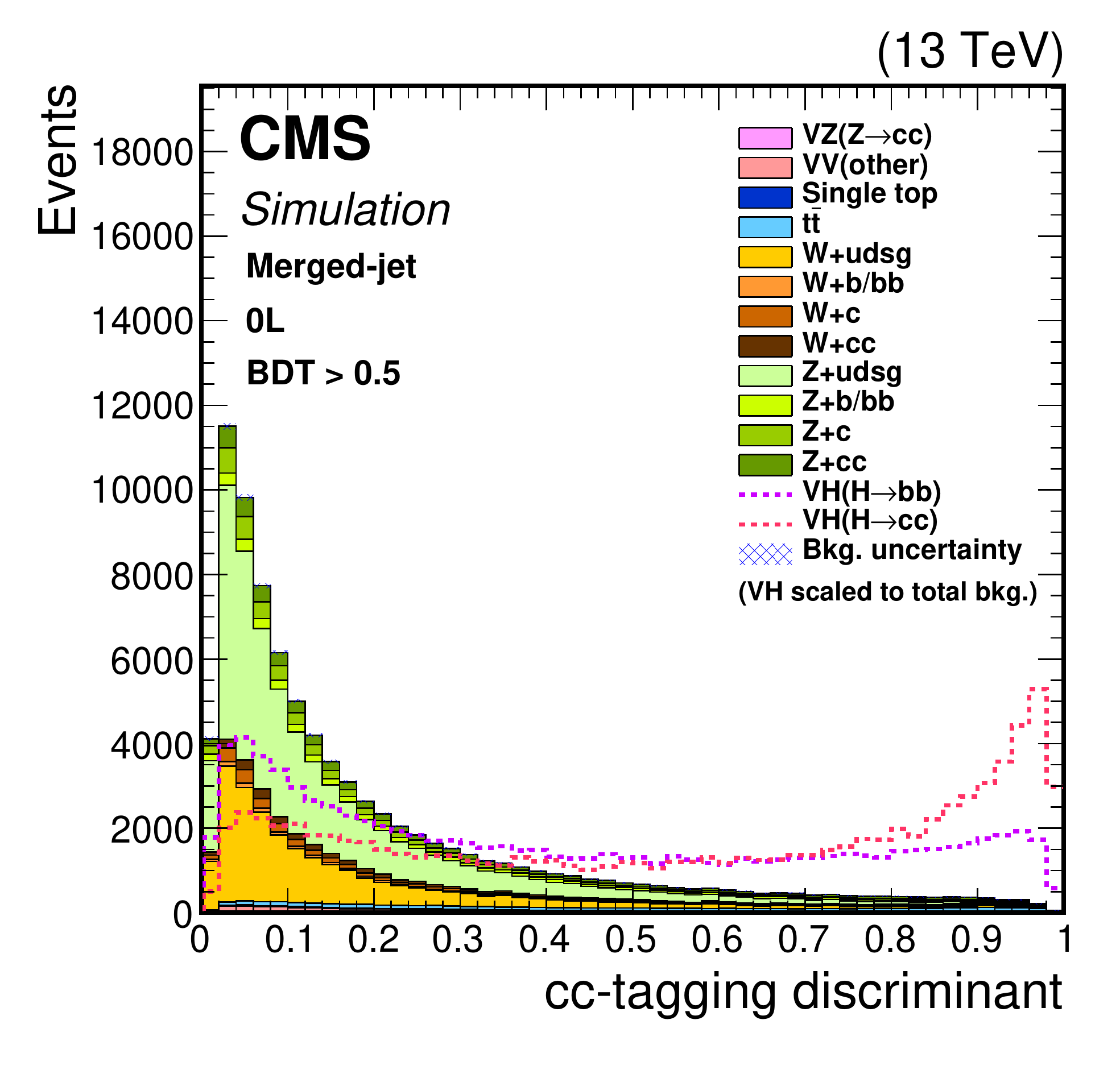}
\includegraphics[width=0.4\textwidth]{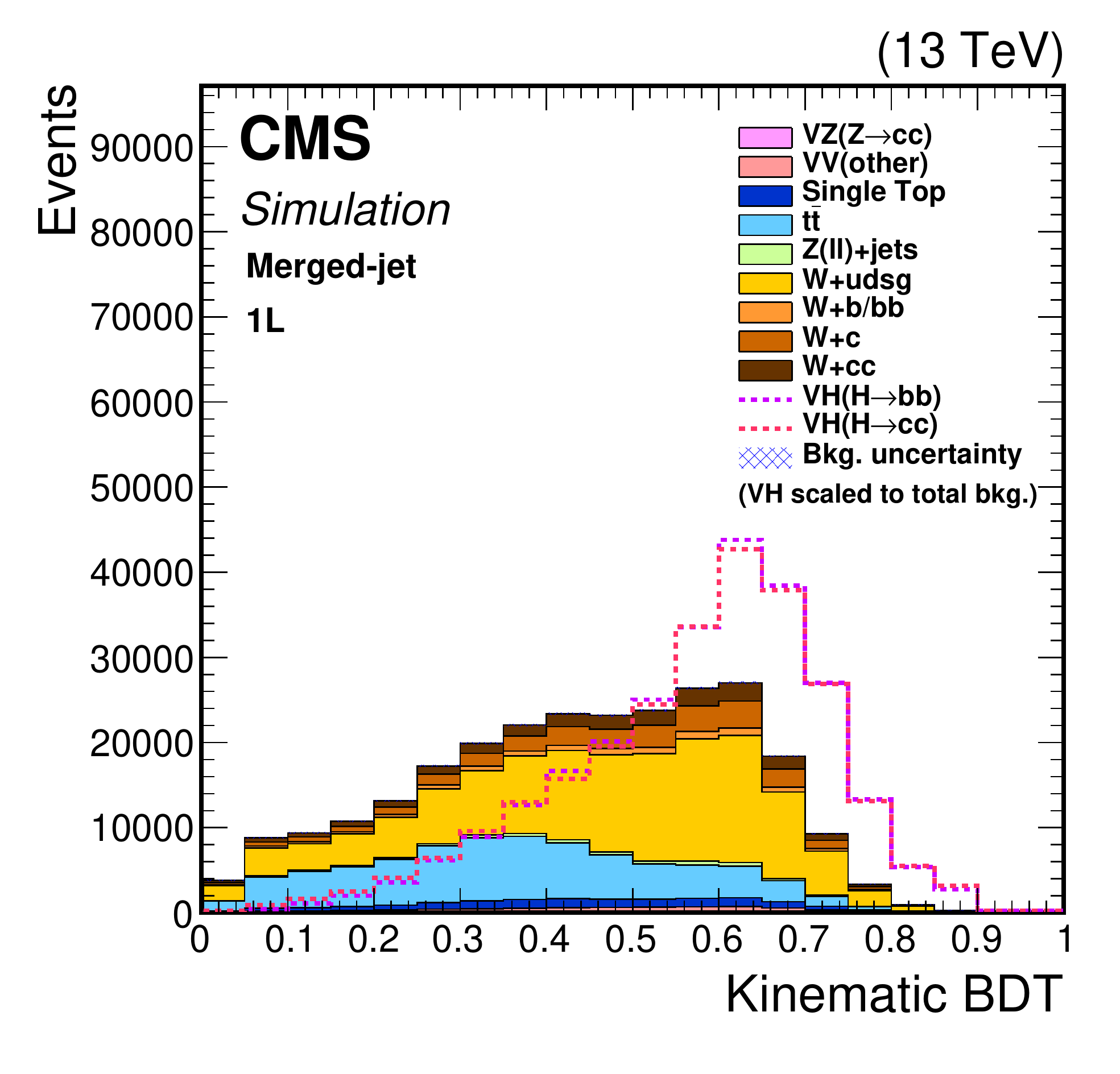}
\includegraphics[width=0.4\textwidth]{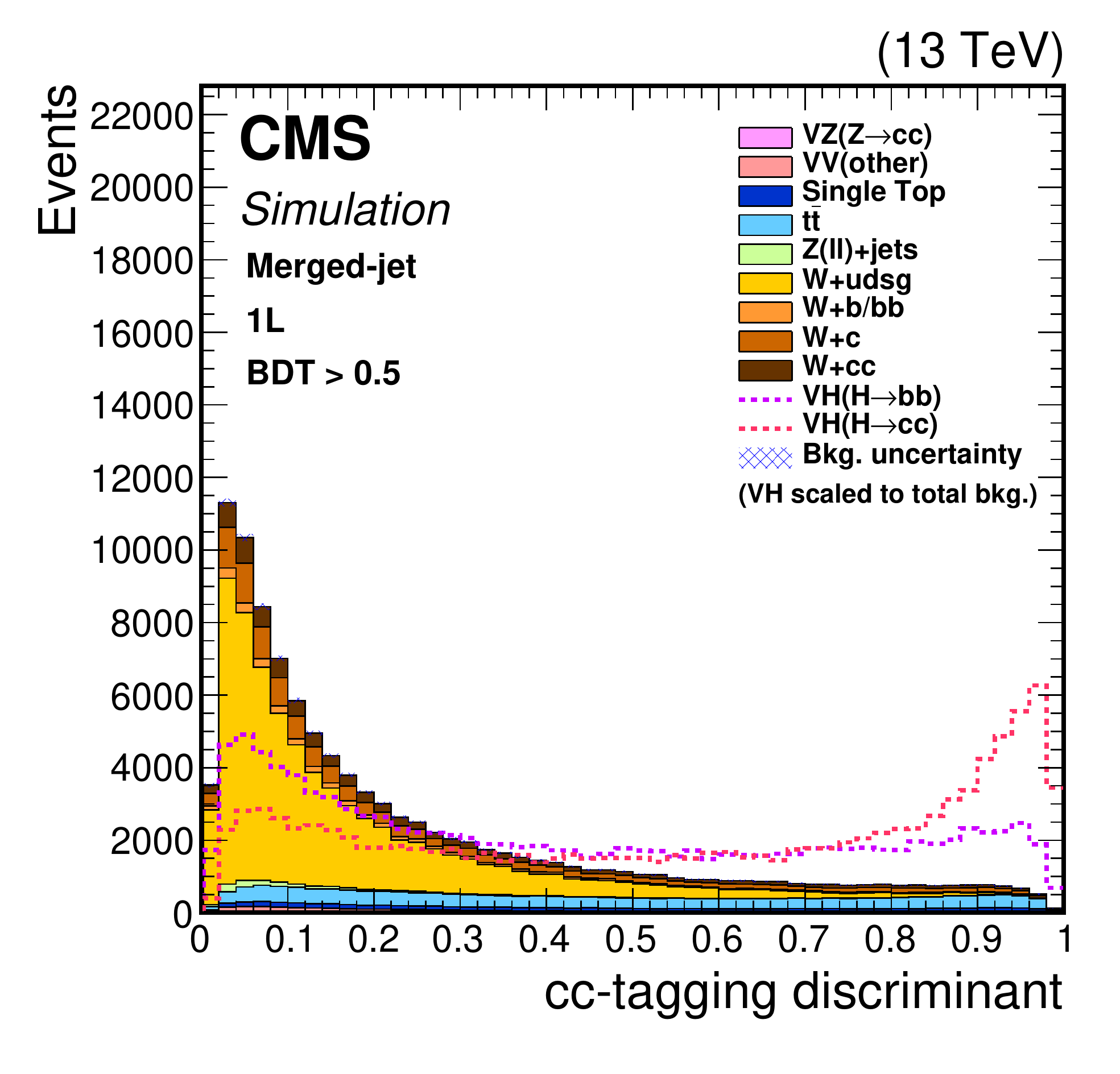}
\includegraphics[width=0.4\textwidth]{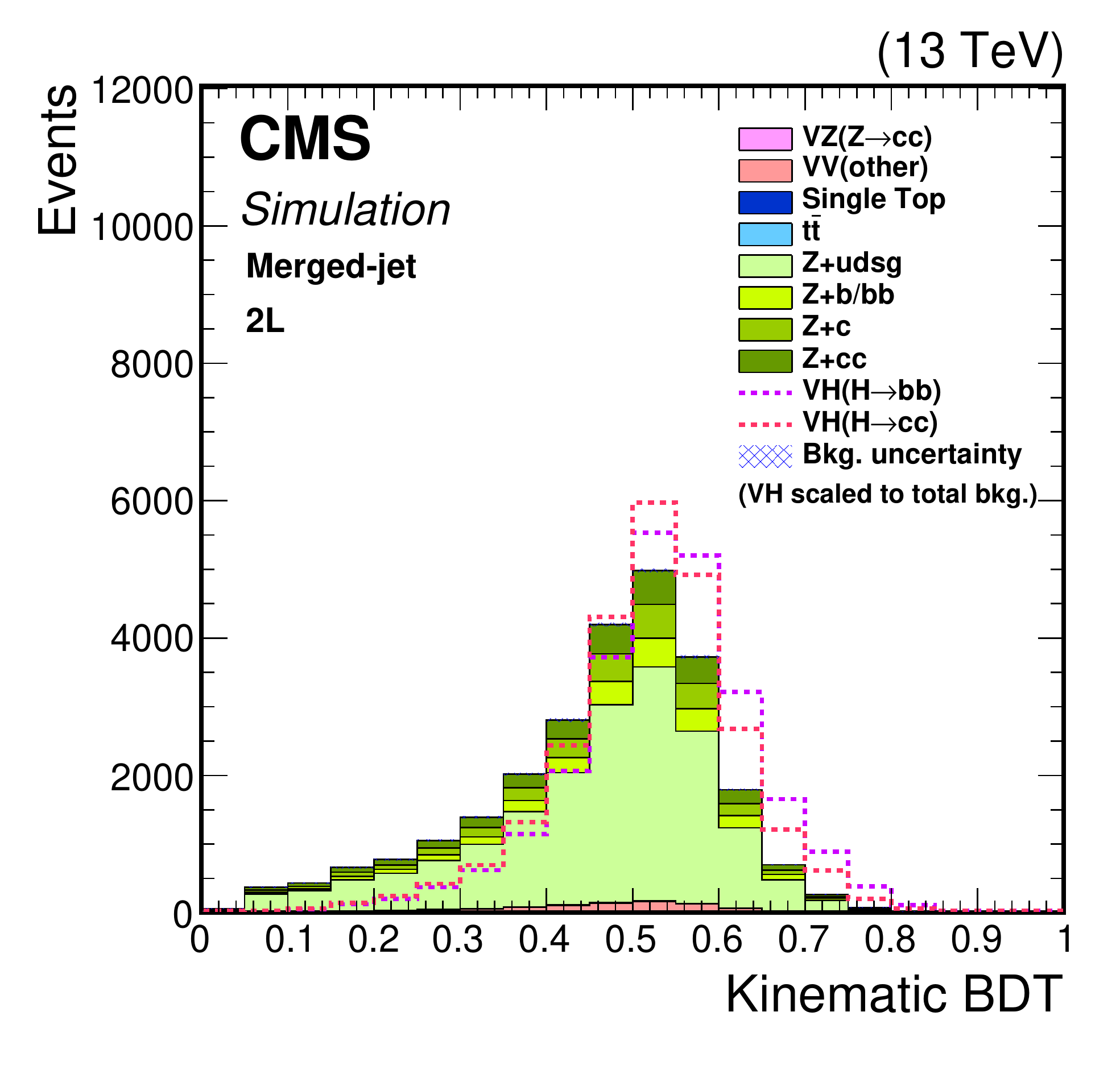}
\includegraphics[width=0.4\textwidth]{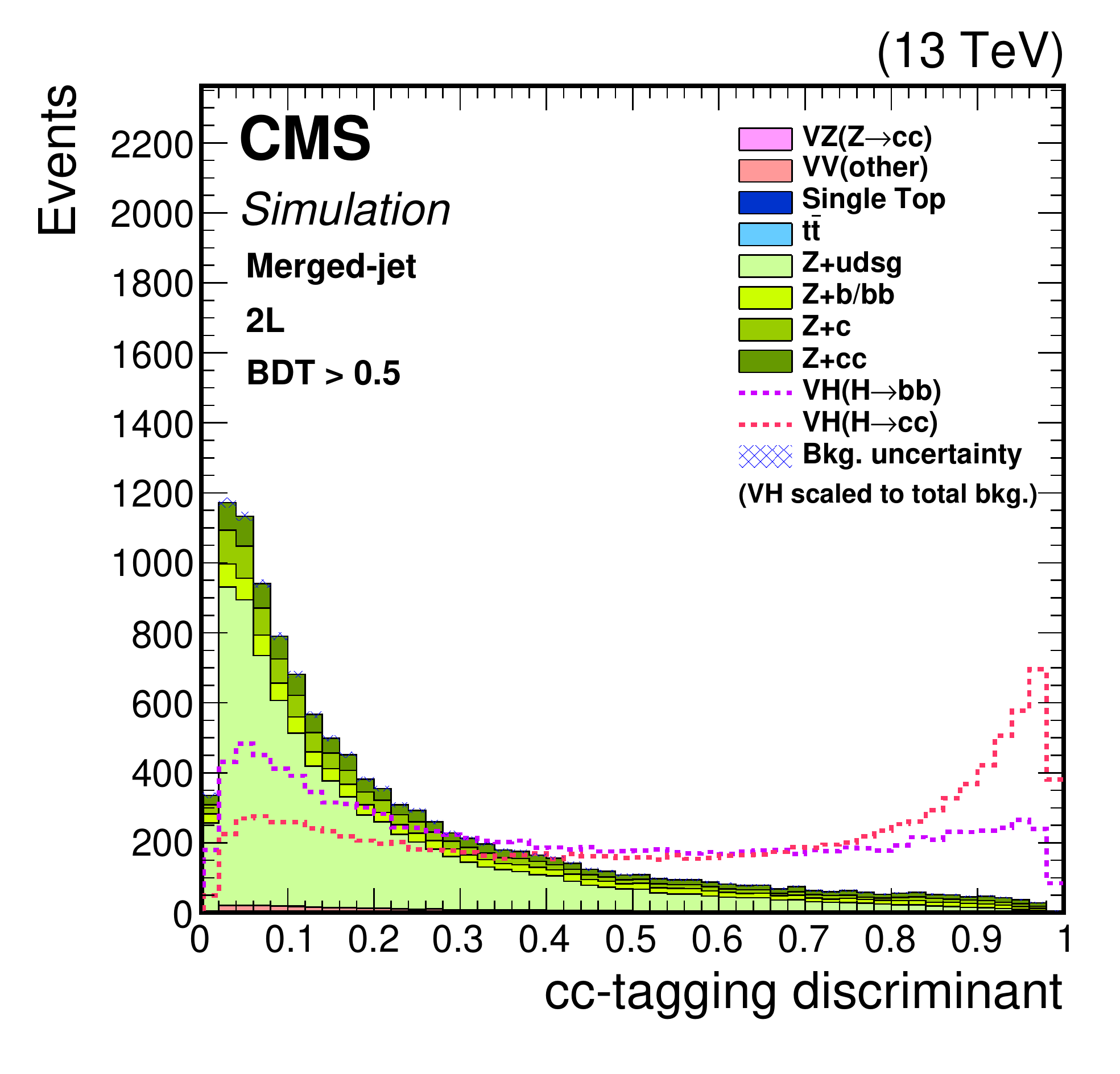}
\caption{The \vhcc signal and background distributions of the kinematic BDT output (left), and the $\ccbar$ discriminant in events with BDT values greater than 0.5 (right), in the 0L (upper), 1L (middle) and 2L (lower) channels. The \vhcc signal is normalised to the sum of all backgrounds. The \vhbb\ contribution, similarly normalised, is also shown.}
\label{fig:2l-kinBDT-DeepAK15}
\end{figure}

The BDT distributions of the three channels for events passing the above selection are shown in Fig.~\ref{fig:2l-kinBDT-DeepAK15} (left) for the \vhcc signal and the background processes.
The discrimination power of the BDT depends on the channel. An improved discrimination power is obtained in the 2L and 1L channels compared
to the 0L channel. In particular, in the 1L channel, improvement is achieved thanks to the presence of the charged lepton and \met, which are then used
for the training of the BDT to provide additional handles to suppress the background.
For all channels, events with BDT values greater than 0.5 define the signal region. The value of 0.5 was obtained in an optimisation for the 1L channel, however further tuning of this value for the 0L and 2L channels has a small impact on the sensitivity. Figure~\ref{fig:2l-kinBDT-DeepAK15} (right) shows the distributions of the $\ccbar$ discriminant in the three channels in the signal region for the \VH signal and the background processes. Good separation is observed between signal and background. The performance of the $\ccbar$ discriminant degrades with the presence of \bq\ quarks, as is the case for \ttbar\ events, for instance.
The signal regions of the merged-jet topology analysis are finally defined requiring the \larger\ jet to pass one of the three working points of the $\ccbar$ discriminant mentioned above.

Dedicated control regions, each enriched in a specific background process, are defined to aid the background estimation in each channel.
Two types of control regions are defined: the ``low-BDT'' control region consisting of events with BDT value $<$0.5, which is enriched in \PV{}+jets background,
and the high-\nsmallr\ control region, defined by inverting the selection on the number of \smallr\ jets to yield a high-purity \ttbar sample.
The latter is not used for the 2-lepton channel since the \ttbar contribution is small in this channel.
In both control regions, events are required to satisfy the same $\ccbar$ tagging discriminant criteria as applied in the signal regions in order
to probe events with a similar flavour composition.
This allows the efficiency of the above mentioned selection to be estimated directly from the data without further corrections being required,
as verified in studies with simulated events and events in data validation regions orthogonal to the control and signal regions.

The low-BDT and the high-\nsmallr\ control regions, together with the signal regions, are included in the maximum likelihood fit to correct for any
difference between data and simulation in the production rate of the \PV{}+jets and \ttbar processes in the phase space selected by this analysis.
Parameters used to separately scale the overall normalisation of the \PW+jets, \PZ+jets, and \ttbar background processes are allowed to float freely in the fit.
These parameters scale the background rate in the same way in both the control and the signal regions.
The parameters are defined separately for each channel, with the exception that the same scale factor is assumed for the \PW+jets process in the 0L and 1L channels. Any potential difference in the $\ccbar$ tagging efficiency between data and simulation is also taken into account in the measured simulation-to-data scale factors.

The merged-jet topology analysis has been validated in two data samples that are completely independent of the control and signal regions.
The first validation sample consists of events with $\ccbar$ tagging discriminant values below those used in the control and signal regions,
and the second validation sample consists of events with lower $\pt(\PH)$. Both validation regions count much more data than the signal
regions and hence can be helpful in identifying potential issues in the analysis. The outcome of these studies is in good agreement with the SM expectation.

\section{Systematic uncertainties}
\label{sec:systematics}
Systematic effects can impact the shape of the BDT discriminant distribution for both analysis topologies,
as well as the \PH candidate mass in the merged-jet topology and the distributions of the charm tagger variables in the resolved-jet topology analysis.
The dominant uncertainties are associated with the normalisation of the background
from the data control regions and the limited size of the dataset. The values of the rate parameters associated with the background normalisations range from $\sim0.65$ to $\sim2.4$ with uncertainties in the range of 10\% to 35\%. Additional systematic effects are related to the jet energy
scale and resolution, which are treated as correlated between the large- and \smallr\ jets.
The efficiency to reconstruct and identify electrons and muons is measured in events with \PZ bosons decaying to electrons and muons via the tag and probe method~\cite{Khachatryan:2010xn}.
The uncertainties on these efficiencies range from 2 to 4\%.
The efficiency for events to pass the combination of triggers based on the presence of one or two leptons is measured using $\PZ/\gamma^{*}\to\Pe\Pe$ and
$\PZ/\gamma^{*}\to\mu\mu$ events via the tag-and-probe method~\cite{Khachatryan:2010xn}. The uncertainties associated with this measurement typically
range between 1 and 3\%. The efficiency of the \ptmiss\ trigger is parametrised using a single-muon dataset and the uncertainties are typically below 2\%.
Theoretical uncertainties related to the cross sections and the \pt\ spectra of the signal and backgrounds are also considered.
In the resolved-jet analysis the systematic uncertainties in PDFs, and in the renormalisation and factorisation scales are treated as uncorrelated
among the four flavour classes considered in the \PV{}+jets processes, as described in Section~\ref{sec:resolvedanalysis}.
The cross section uncertainties are set to their theoretical predictions, while the systematics associated to the renormalisation and factorisation scales are obtained varying the parameters
relative to these scales to 0.5 and 2.0 of their nominal values and estimating the effect. Lastly, the uncertainties in the \cq\ quark identification are also considered.
The full list of systematic uncertainties is provided in Table~\ref{tab:systematics}.

In Table~\ref{tab:systematicsImpact} the uncertainty sources are grouped into categories and their impact on the fitted signal strength resulting
from the combination of the resolved-jet and merged-jet topology analyses is provided (see Section~\ref{sec:results} for more details).
The uncertainty breakdown shows that the search for the \vhcc process is mainly limited by the size of the available dataset:
the related uncertainty accounts for more than 85\% of the total uncertainty in the fitted $\mu$.
The statistical uncertainties include contributions from the limited number of events in the available dataset and the background normalisations extracted from the control regions.
The main sources of systematic uncertainties come from the charm tagging efficiencies and the modelling of the simulated physics processes, representing $\sim28\%$ and $\sim25\%$
of the total uncertainty, respectively. The uncertainties in the theory prediction, which include uncertainties in the cross sections, \pt spectra, PDFs, renormalisation and
factorisation scales, also play a considerable role and represent approximately 30\% of the total uncertainty in $\mu$.

\begin{table*}[htbp]
\topcaption{Summary of the systematic uncertainties for each channel. The shape uncertainties refer to systematic uncertainties that affect both the shape of the distribution being
fitted as well as their normalisation. Uncertainties in the lepton identification and trigger efficiencies are treated as a normalisation uncertainty in the resolved-jet topology
analysis and as a shape uncertainty in the merged-jet topology analysis.}
\label{tab:systematics}
\centering
{
\begin{tabular}{lcccc}
\hline
Source & Type & 0-lepton & 1-lepton & 2-lepton \\
\hline
Simulated sample event count                                               & shape   & \xmark & \xmark & \xmark \\
\PQc tagging efficiency                                                    & shape   & \xmark & \xmark & \xmark \\
Top quark \pt reweighting                                                  & shape   & \xmark & \xmark & \xmark \\
\ptV reweighting                                                           & shape   & \xmark & \xmark & \xmark \\
\VH: \ptV NLO EW correction                                                & shape   & \xmark & \xmark & \xmark \\
Jet energy scale                                                           & shape   & \xmark & \xmark & \xmark \\
Jet energy resolution                                                      & shape   & \xmark & \xmark & \xmark \\
\ptmiss\ trigger efficiency                                                & rate    & $2\%$ & \NA    & \NA    \\
Lepton trigger efficiency                                                  & shape (rate) & \NA      & \xmark & \xmark \\
Lepton identification efficiency                                           & shape (rate)  & \NA       & \xmark & \xmark \\
\ptmiss\ unclustered energy                                                & shape   & \xmark & \xmark & \NA    \\
Pileup reweighting                                                         & shape   & \xmark & \xmark & \xmark \\
Integrated luminosity                                                      & rate    & $2.5\%$ & $2.5\%$  & $2.5\%$  \\
PDF                                                                        & shape   & \xmark & \xmark & \xmark \\
Renormalisation and factorisation scales                                   & shape   & \xmark & \xmark & \xmark \\
Single top cross section                                                   & rate    & $15\%$ & $15\%$ & $15\%$ \\
Diboson cross section                                                      & rate    & $10\%$ & $10\%$ & $10\%$ \\
\VH: cross section (PDF)                                                 & rate    & $1.6\%-2.4\%$ & $1.9\%$ & $1.6\%-2.4\%$ \\
\VH: cross section (scale)                                               & rate    & $19\%-25\%$ & $0.5\%-0.7\%$ & $19\%-25\%$ \\
\hline
\end{tabular}
}
\end{table*}

\begin{table*}[htbp]
\caption{Summary of the impact of the statistical and systematic uncertainties on the signal strength modifier for combined analysis of the resolved-jet and merged-jet topologies.}
\label{tab:systematicsImpact}
\centering
{
\begin{tabular}{lrr}
\hline
Uncertainty source                          &   \multicolumn{2}{c}{$\Delta\mu \mid \mu=37$}\\
\hline

Statistical                         &  $+17.3$ &   $-17.1$ \\
\hspace{5mm}    Background normalisations               &  $+10.1$ &  $-10.2$ \\

Experimental                        &  $+7.6$ &   $-8.2$ \\
\hspace{5mm}    Charm tagging efficiencies              &  $+5.6$ &   $-4.8$ \\
\hspace{5mm}    Simulation modeling                     &  $+4.2$ &   $-5.1$ \\
\hspace{5mm}    Jet energy scale and resolution         &  $+2.4$ &   $-2.8$ \\
\hspace{5mm}    Lepton identification efficiencies      &  $+0.4$ &   $-1.8$ \\
\hspace{5mm}    Luminosity                              &  $+1.6$ &   $-1.7$ \\
\hspace{5mm}    Statistics of the simulated samples     &  $+0.5$ &   $-1.9$ \\

Theory                              &  $+6.5$ &   $-4.6$ \\
\hspace{5mm}    Signal                                  &  $+5.0$ &   $-2.5$ \\
\hspace{5mm}    Backgrounds                             &  $+4.3$ &   $-3.9$ \\

Total                               & $+20.0$  &   $-19.5$ \\

\hline
\end{tabular}
}
\end{table*}

\section{Results}
\label{sec:results}
The signal extraction strategy is based on a binned likelihood fit to the data,
with the signal and control regions fitted simultaneously.
Separately fitting electron and muon events in the 1L and 2L channels is driven by the fact that muons have a significantly smaller misreconstruction
rate and greater signal sensitivity. In addition, the trigger, identification and isolation scale factors are different because electrons and muons are reconstructed differently with the CMS detector.
The upper limit (UL) on the signal strength $\mu$ for SM production of \vhcc\ is extracted at 95\%
\CL based on a modified frequentist approach~\cite{CLS1,CLS2} under the asymptotic approximation for the profile likelihood test statistic~\cite{Cowan:2010js,CMS-NOTE-2011-005}.
Both analyses are validated by measuring the products of the \VZ production cross section and the branching fraction of \PZ to charm quark-antiquark pair, $\mathcal{B}\left(\PZ \to \cq\cqbar \right)$.
The systematic uncertainties are incorporated in the fit as constrained parameters of the likelihood function. The cross section of the \vhbb process is set to its SM prediction for the \PH boson mass of 125\GeV.
The result has only a weak dependence on the assumed \vhbb rate. Indeed, on average, the energy carried by neutrinos is higher for the \vhbb than for the \vhcc process.
This leads to distinguishably different contributions to the final fitted distribution. Varying the \vhbb rate by 100\% results in less than a 5\% change in the expected sensitivity.

The results obtained in the resolved-jet and merged-jet topology analyses independently, i.e.,
exploiting larger regions of the full phase space prior to defining disjoint data samples for the combination of results, are described in Sections \ref{sec:results_reso} and \ref{sec:results_boost}.
As described in Sections~\ref{sec:resolved_an} and \ref{sec:boosted_an}, in the merged-jet topology analysis the phase-space considered is bounded from below by $\Vpt>200\GeV$, while for the resolved-jet topology analysis the lower bound is set by the \Vpt\ thresholds of 50, 100, and 170\GeV in the 2L, 1L and 0L channels, respectively. Neither of the analyses has an upper limit on \Vpt.
The two analyses are then combined for the final result, presented in Section \ref{sec:combination},
after making them statistically independent via a selection on \Vpt\ to set an upper bound for the resolved-jet topology analysis that is also the lower bound on the merged-jet topology analysis.

\subsection{Resolved-jet topology}
\label{sec:results_reso}
In the resolved-jet topology analysis, the \vhcc\ signal is extracted via a binned likelihood
fit to the BDT output distributions, that is carried out simultaneously with fits to the backgrounds in control regions.
In the LF control regions the fits are for the $\textit{CvsL}_{\text{min}}$ distributions, while in the TT, HF, and CC control regions they are for the $\textit{CvsB}_{\text{min}}$ distributions, as detailed in Section~\ref{sec:ResoStrategy}.

The analysis is first validated by measuring the product of the \VZ production cross section and $\mathcal{B}\left(\PZ \to \cq\cqbar \right)$ normalised to the SM prediction.
A separate BDT is trained for each channel, using \vzcc\ as signal and \vhcc\ as contribution to background with cross section fixed to the SM prediction. The measured signal strength for the \vzcc\ process is $\mu_{\vzcc}=1.35^{+0.94}_{-0.95}$ with an observed (expected) significance of 1.5 $(1.2)$ standard deviations ($\sigma$), respectively.
The results are consistent within uncertainties with the SM expectation.

\begin{figure}[!hp]
  \centering
    \includegraphics[width=0.37\textwidth]{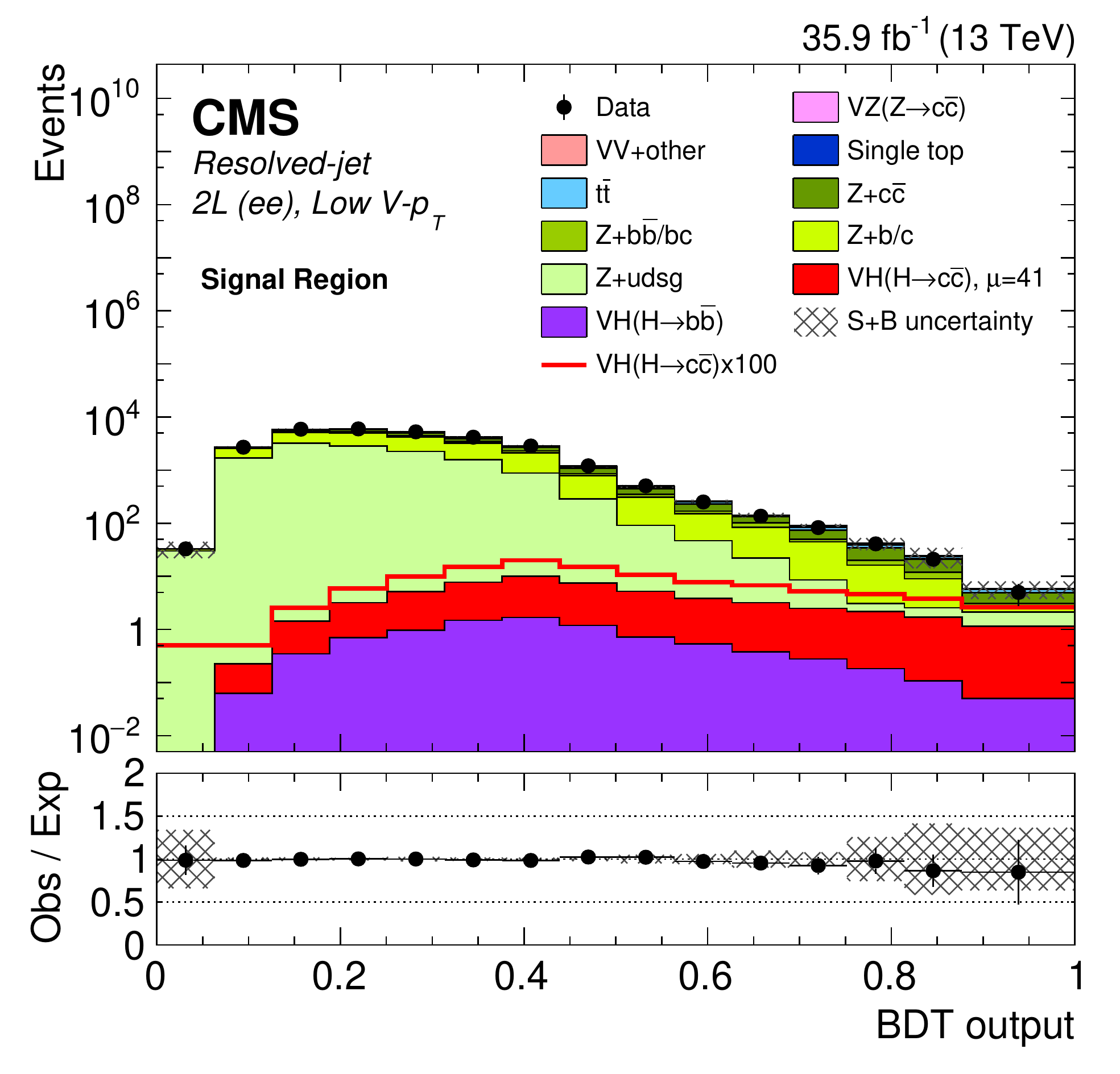}
    \includegraphics[width=0.37\textwidth]{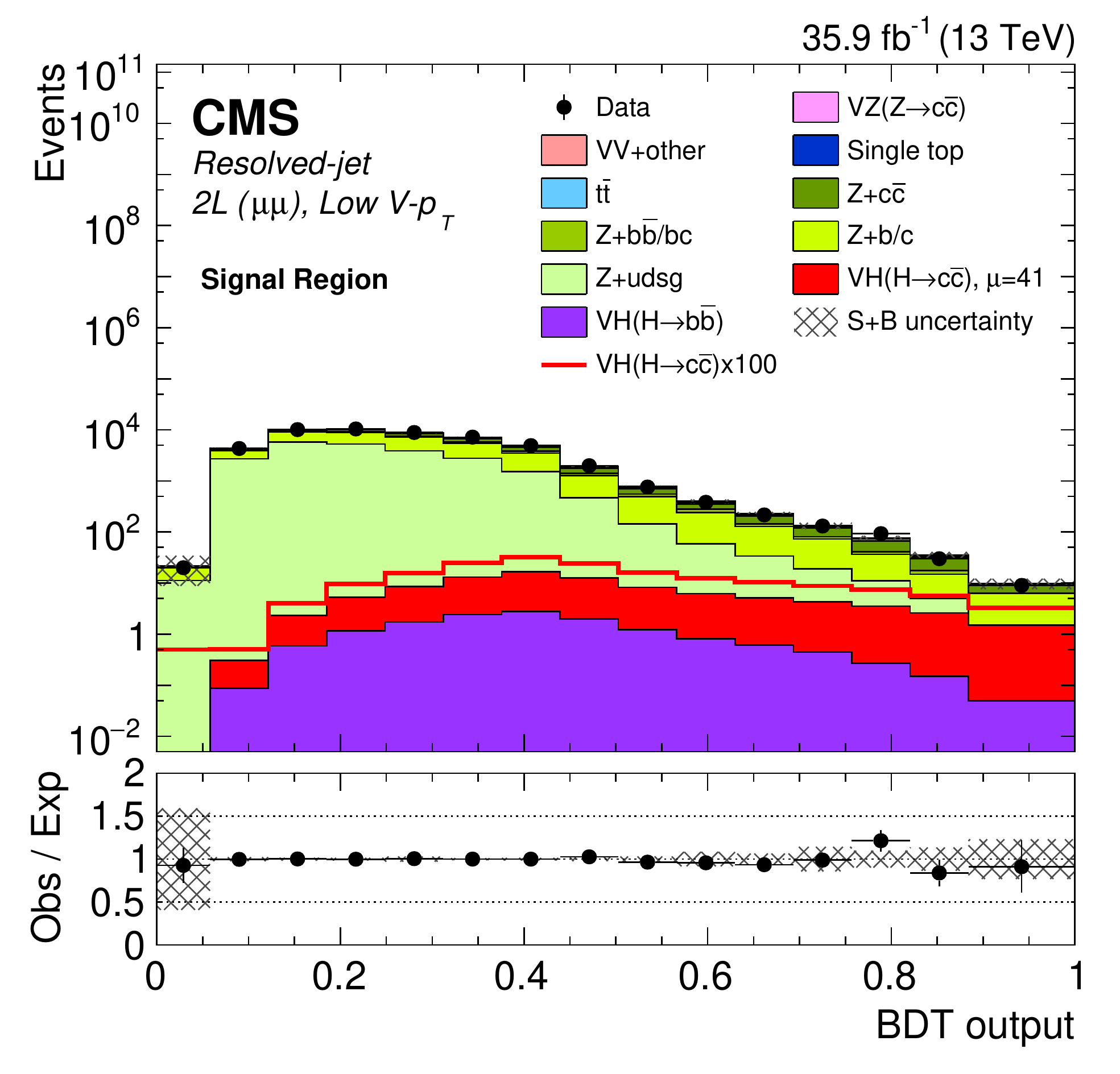}\\
    \includegraphics[width=0.37\textwidth]{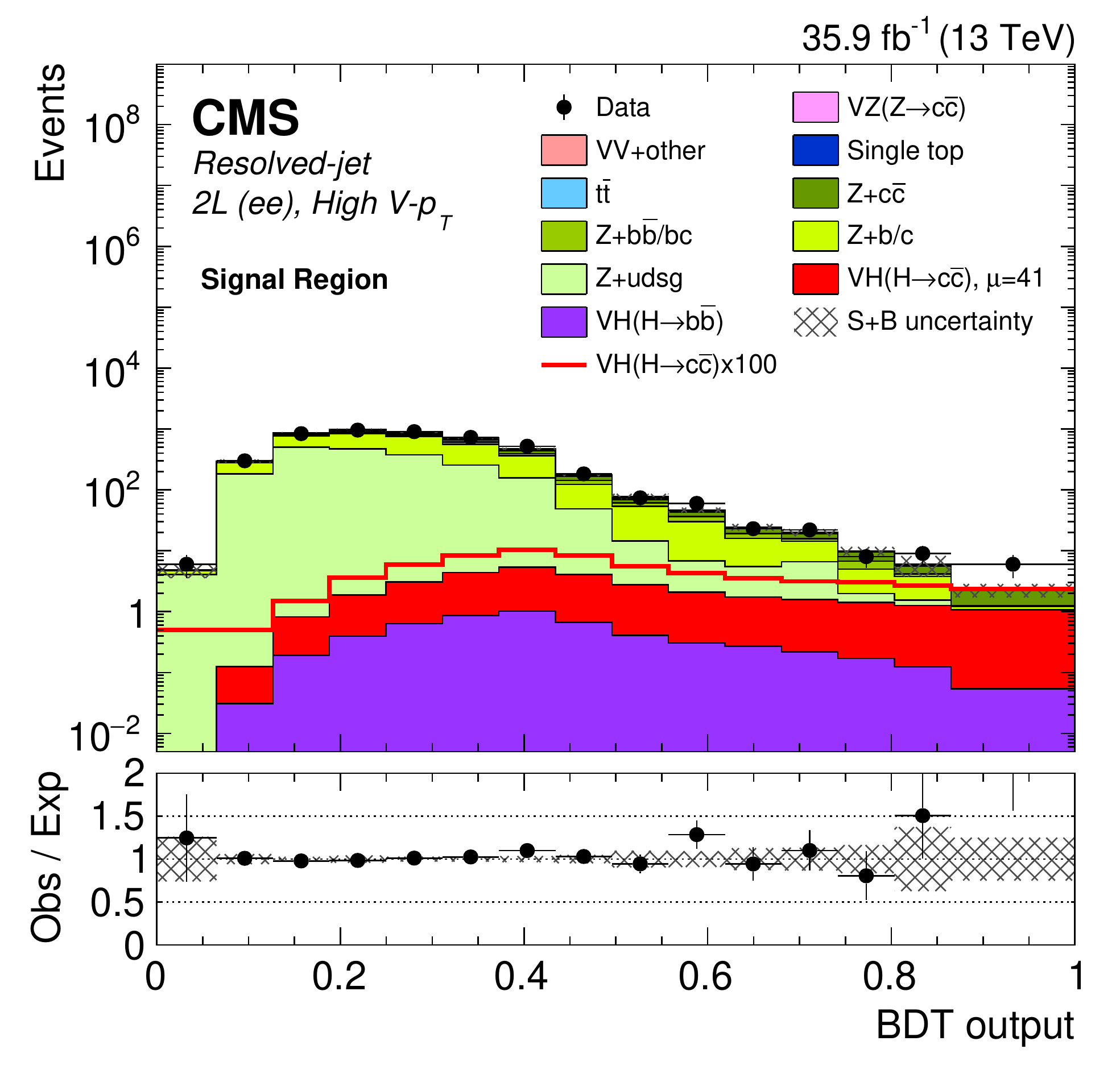}
    \includegraphics[width=0.37\textwidth]{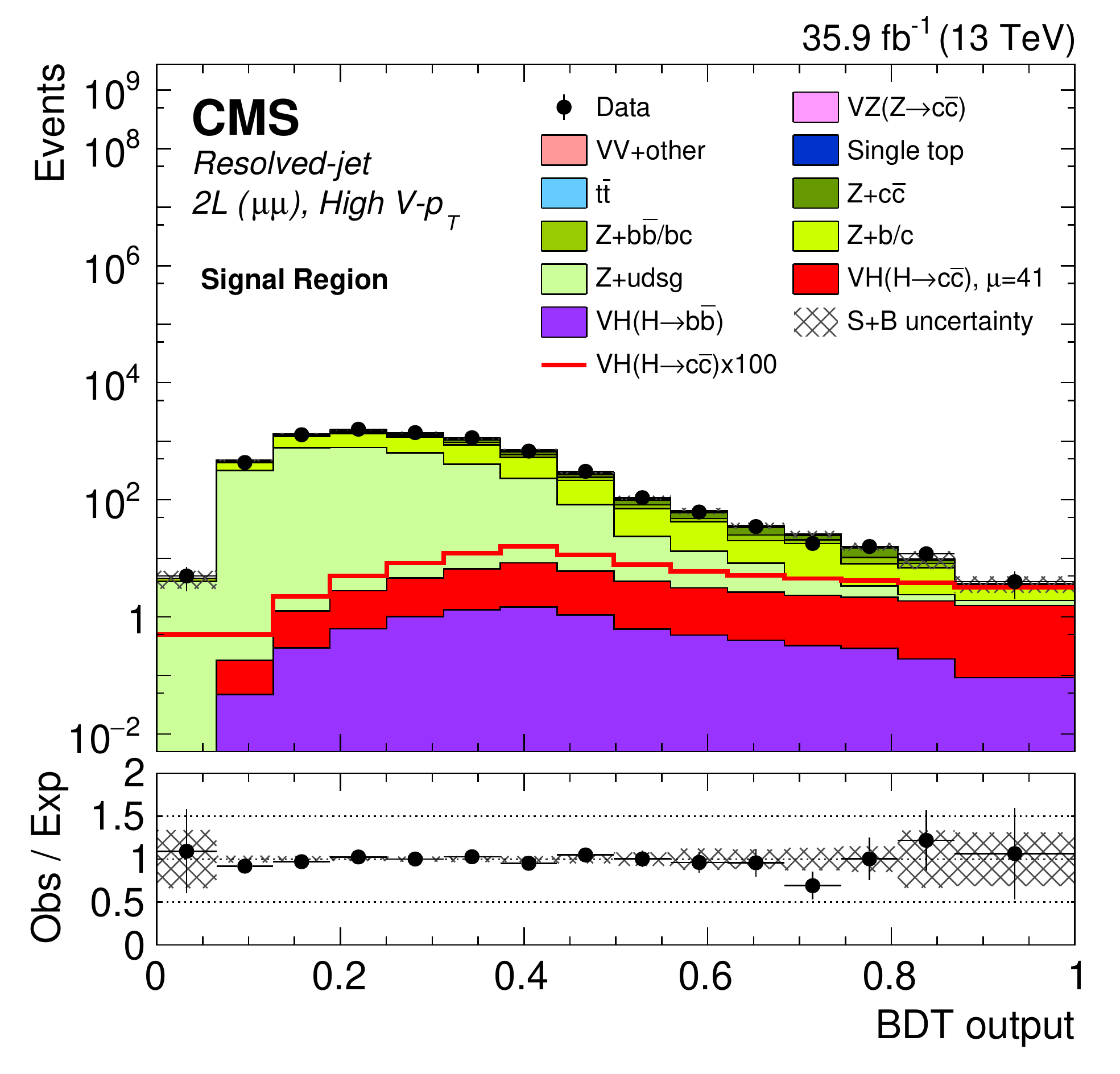}\\
    \includegraphics[width=0.37\textwidth]{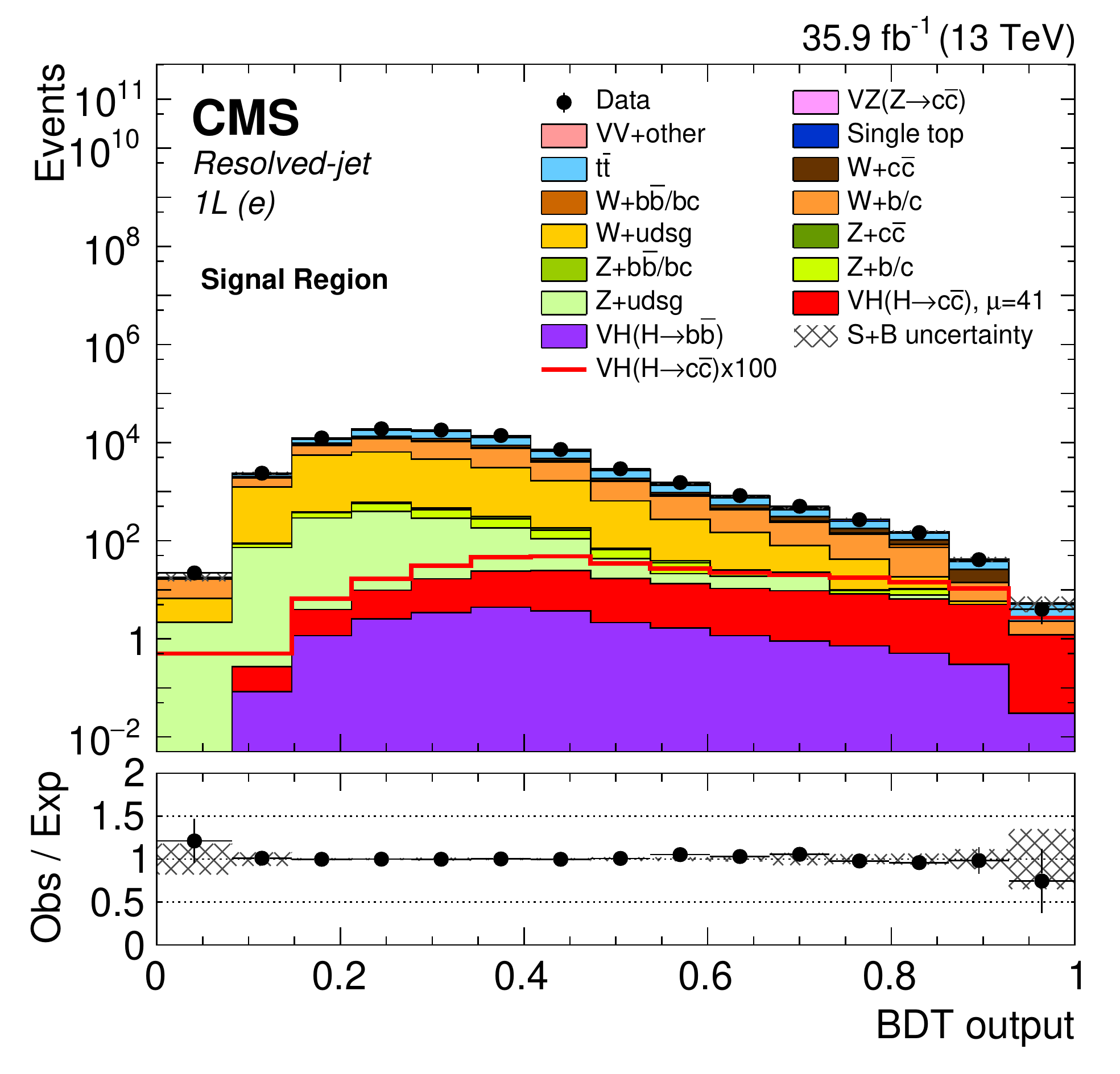}
    \includegraphics[width=0.37\textwidth]{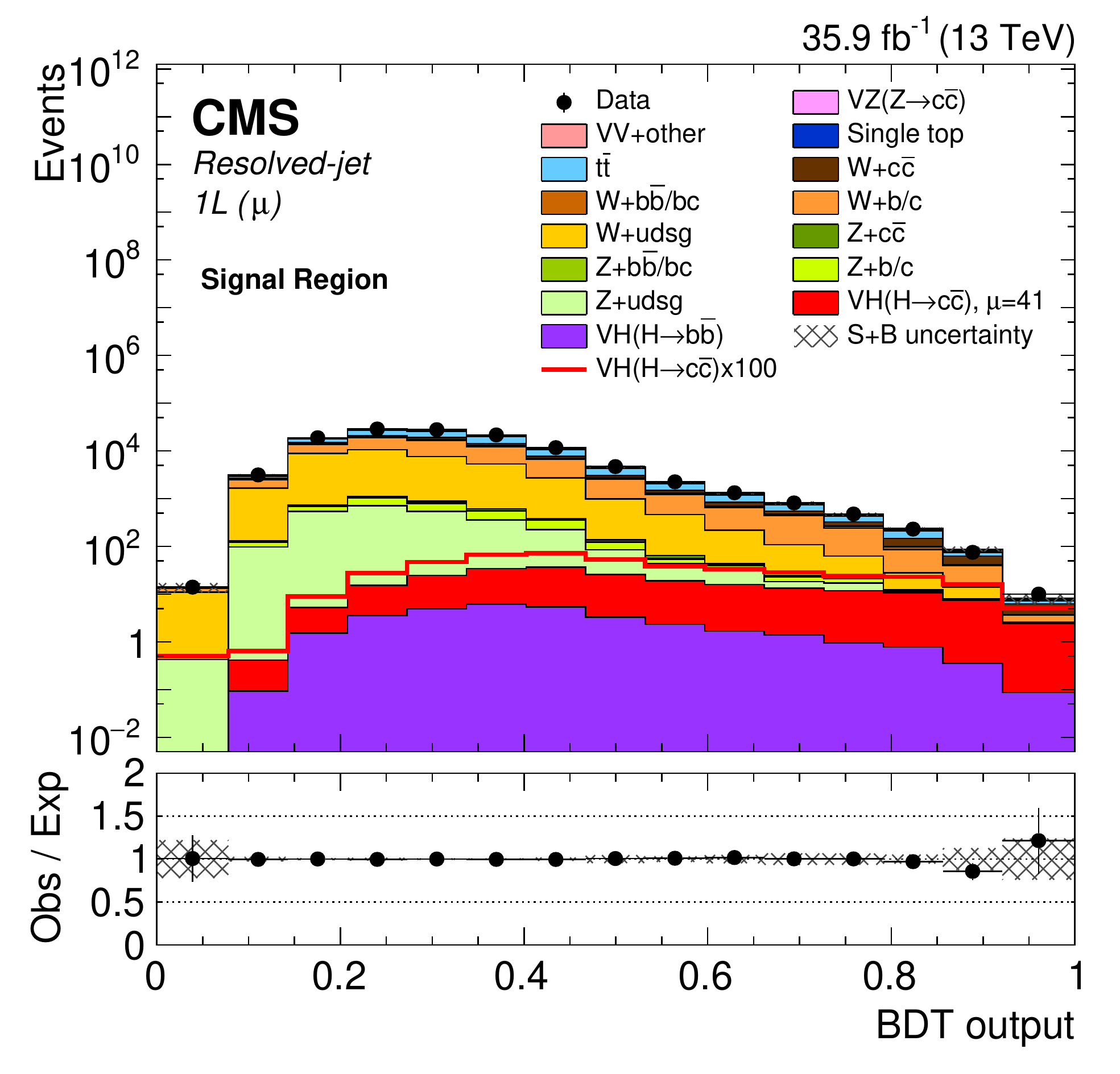}
    \includegraphics[width=0.37\textwidth]{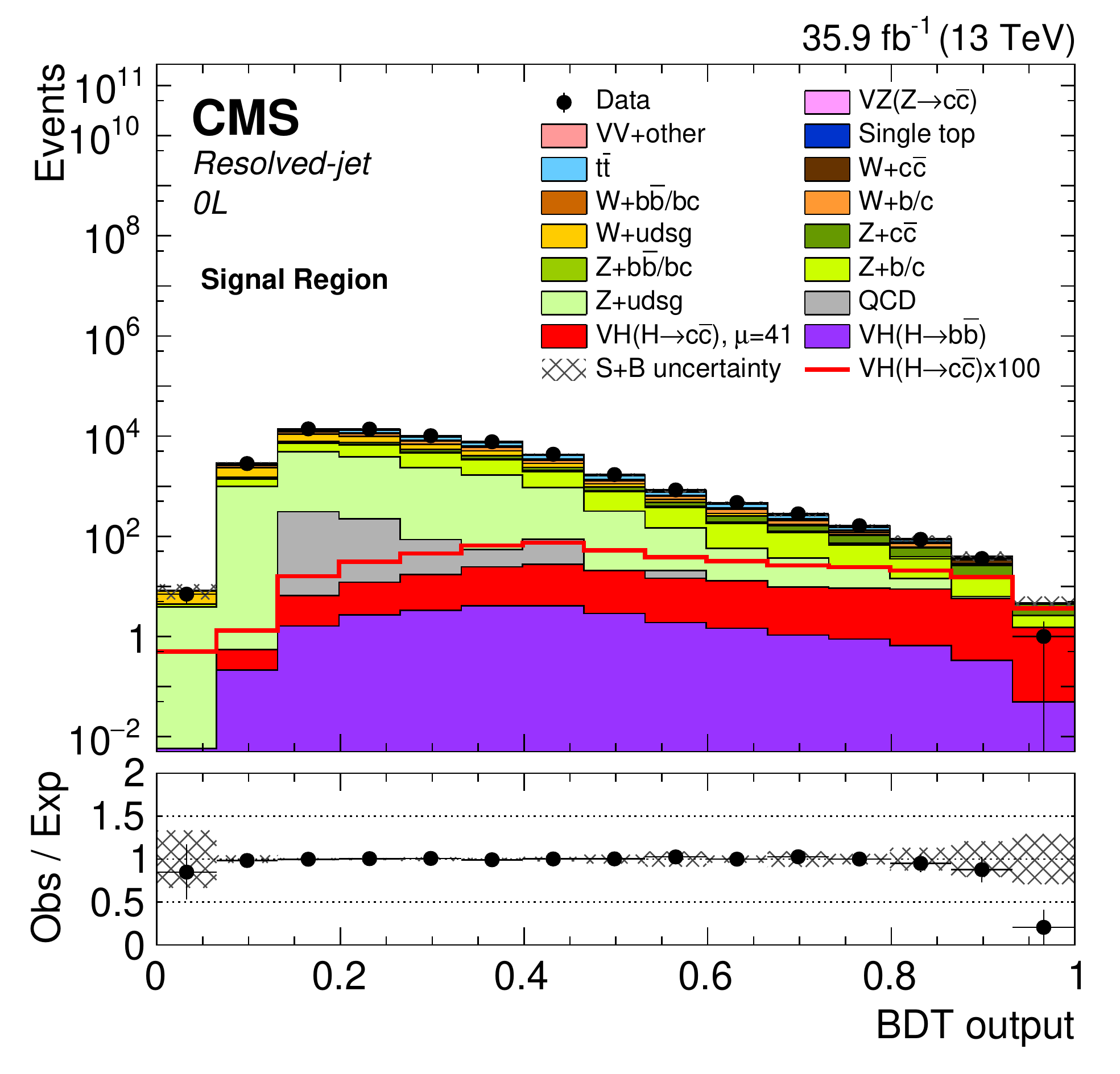}
    \caption{Post-fit distributions of the BDT score in the signal region of the resolved-jet topology analysis for the 2L low-\Vpt, 2L high-\Vpt, 1L, and 0L channels. The plain red histograms represent the signal contribution normalized by the post-fit value of $\mu_{\vhcc}$, while the red line histograms show the expected signal contribution multiplied by a factor 100.  }
    \label{fig:PostFitSR}
\end{figure}

A dedicated BDT is trained for each channel to distinguish the \vhcc\ signal from the backgrounds. Figure~\ref{fig:PostFitSR} displays the BDT distributions in all search channels after the fit to the data. In all plots, the value of each nuisance parameter has been fixed to its best fit value. In general, the BDT distributions in data agree well with the background predictions. The largest excess in the data occurs at large BDT values in the high-\Vpt\ 2L (\Zee) channel with an observed local signal significance of $2.1\,\sigma$.

The observed (expected) UL at 95\% CL on $\mu$ for SM \vhcc\ production is 75 ($38^{+16}_{-11}$), and the measured signal strength is $\mu_{\vhcc}=41^{+20}_{-20}$. The uncertainties in the expected UL correspond to a variation of $\pm1\,\sigma$ in the expected event yields under the background-only hypothesis. The results are consistent with the SM expectations within two standard deviations. This modest deviation is mostly due to the small excess mentioned above. The results for each channel and their combination are shown in Table~\ref{tab:limits-ind}. The most sensitive channel is 2L, whereas the 0L and 1L channels have similar sensitivity.

\subsection{Merged-jet topology}
\label{sec:results_boost}
\begin{figure}[hp]
\centering
\includegraphics[width=0.45\textwidth]{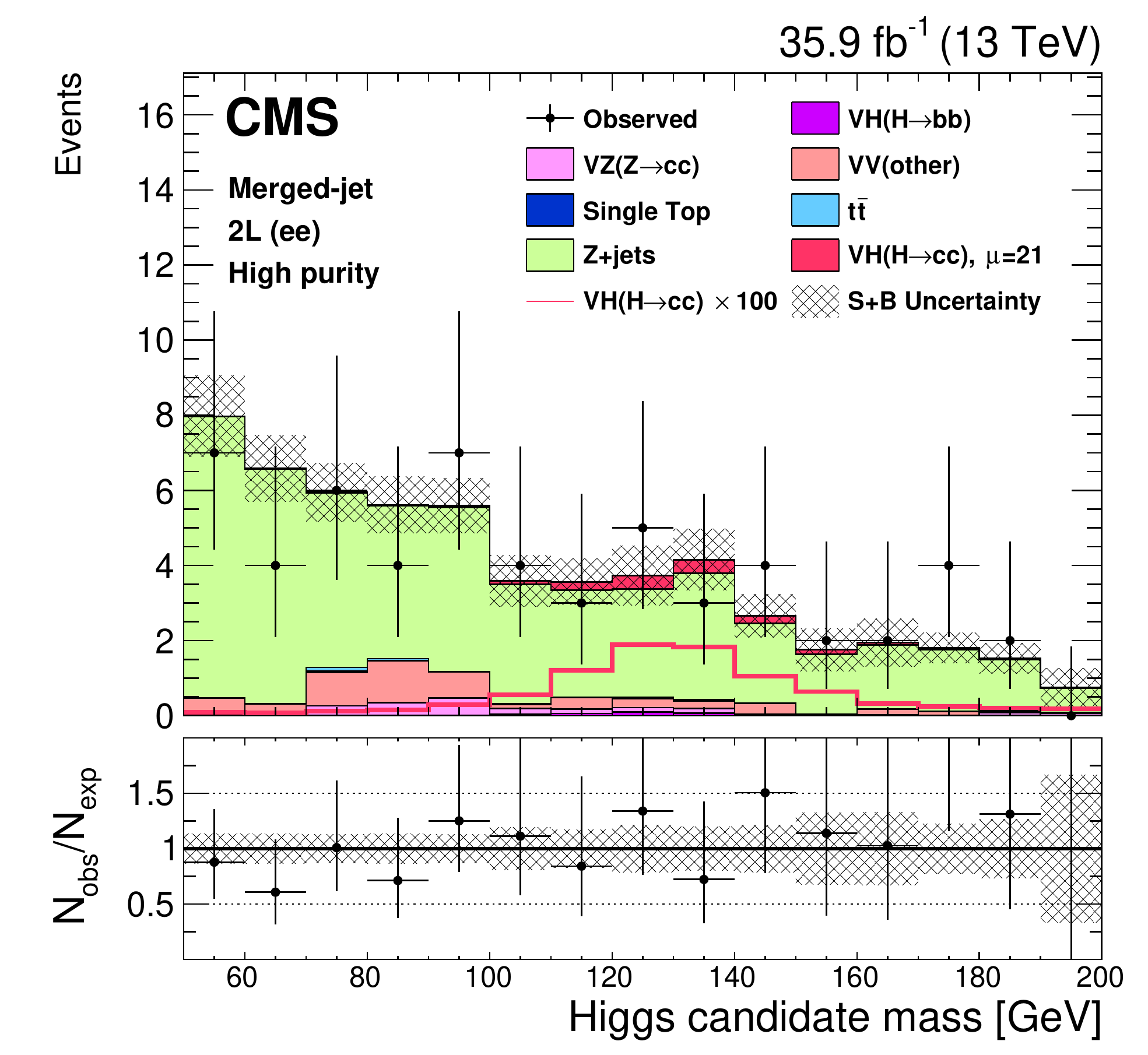}
\includegraphics[width=0.45\textwidth]{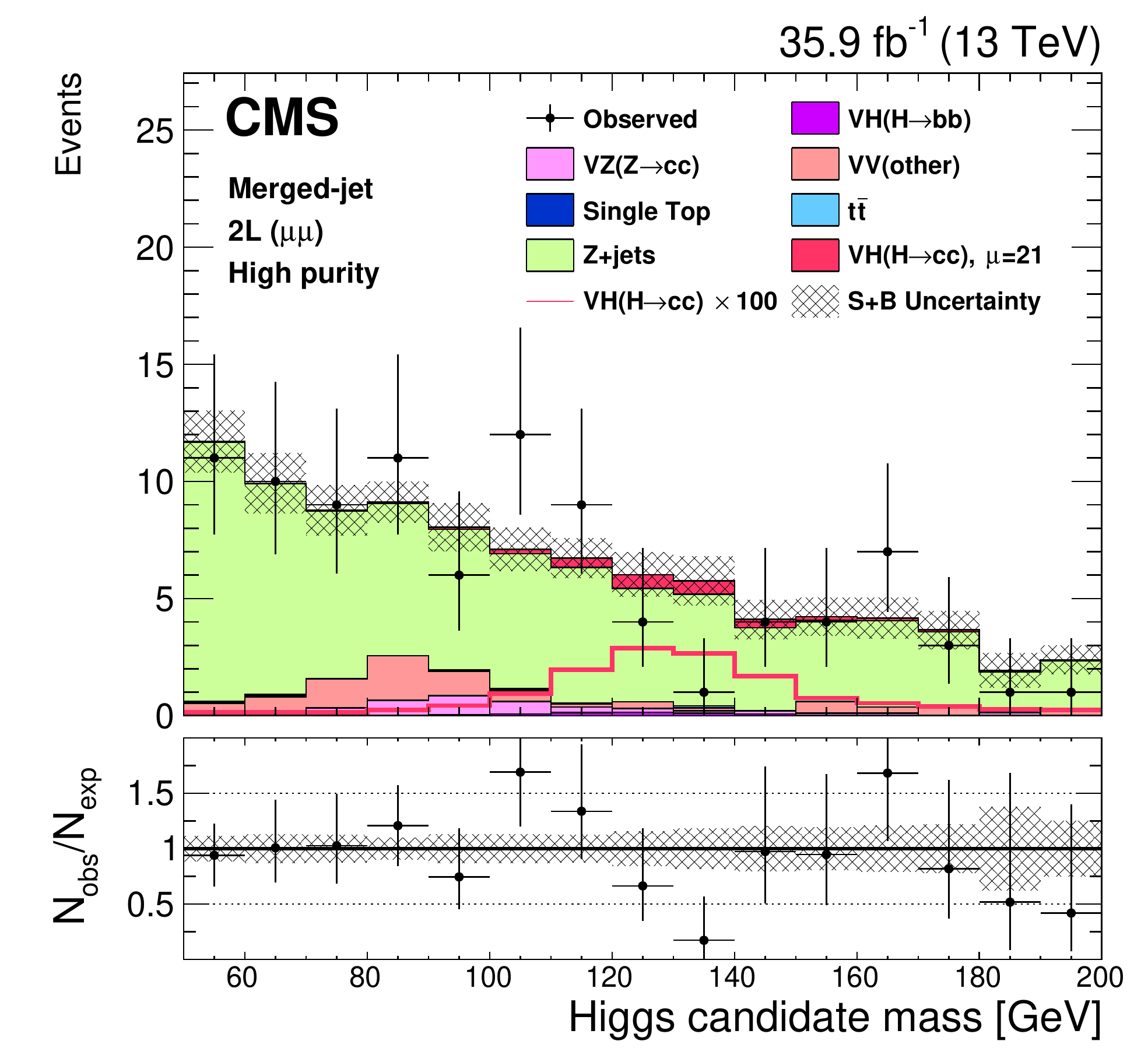} \\
\includegraphics[width=0.45\textwidth]{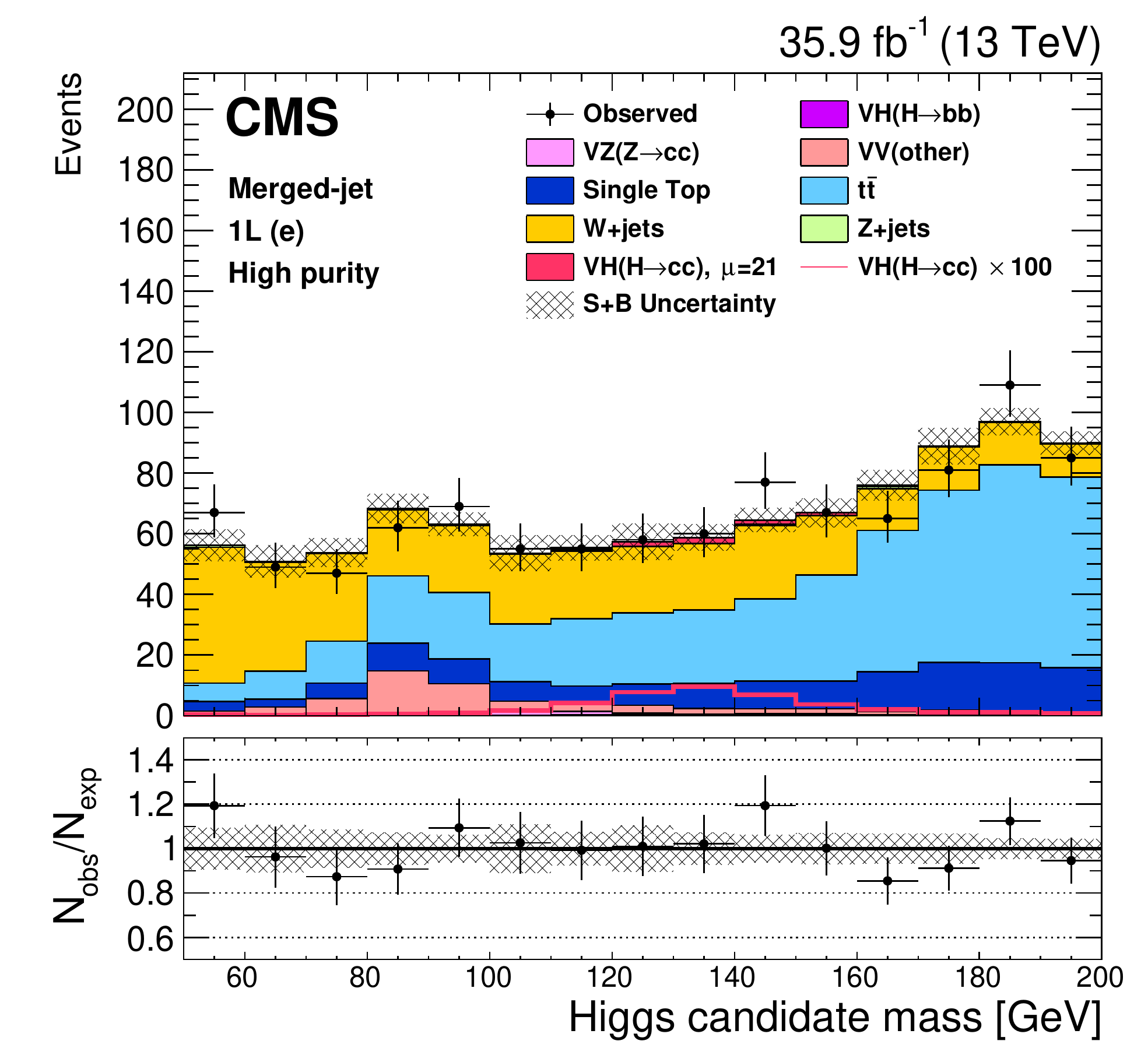}
\includegraphics[width=0.45\textwidth]{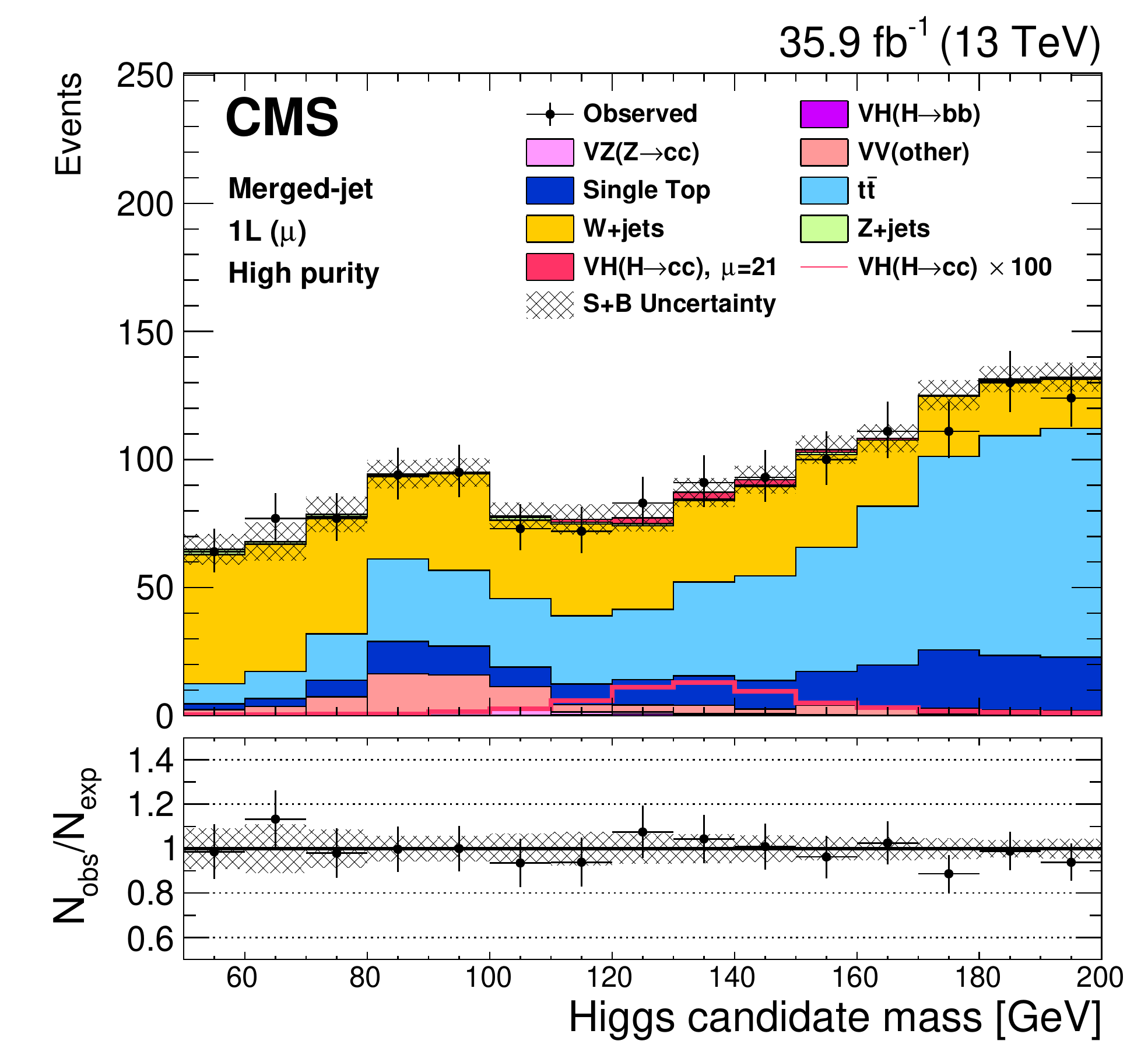}\\
\includegraphics[width=0.45\textwidth]{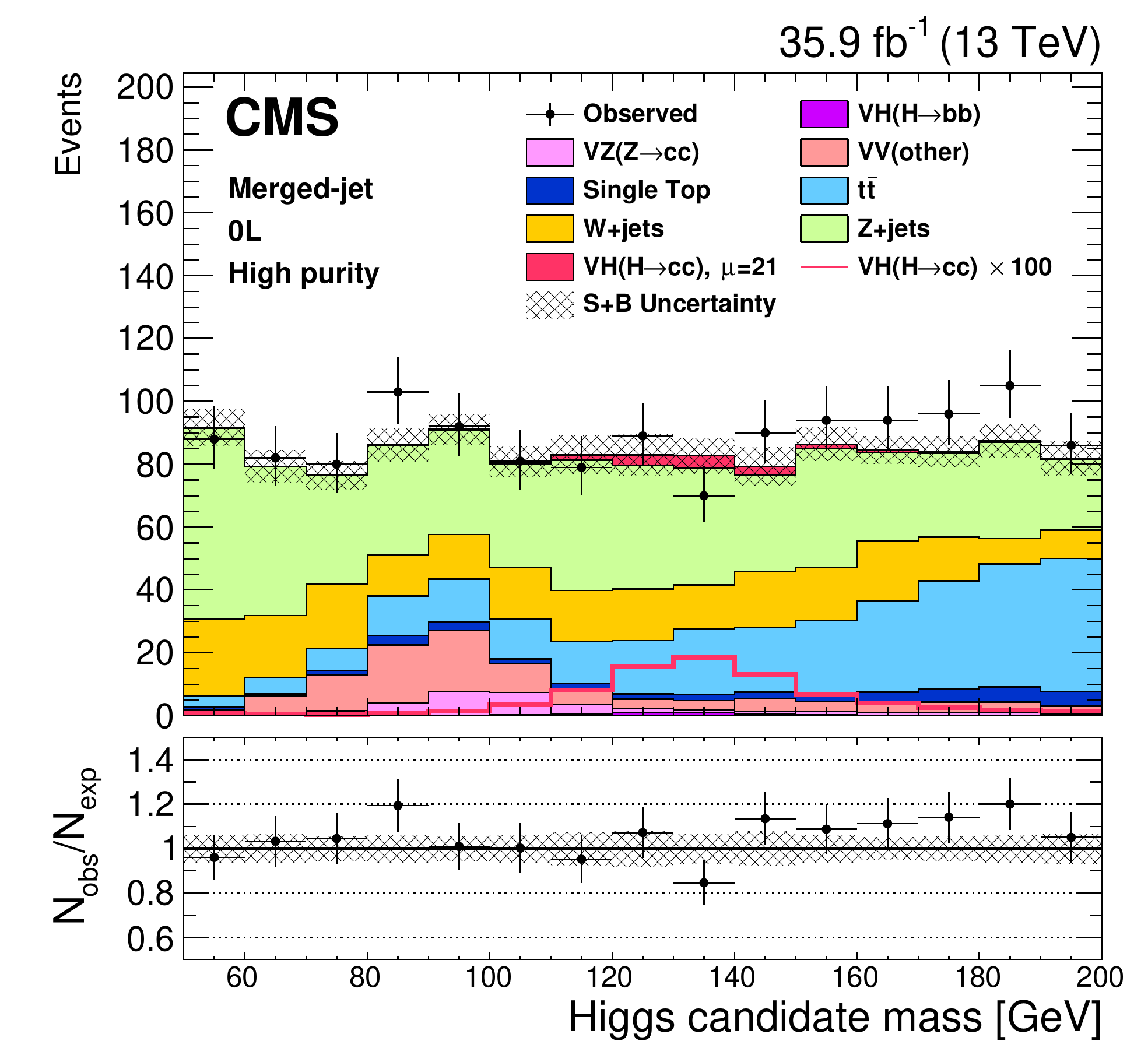}
\caption{The \msd\ distribution of \PH in data and simulation in the merged-jet topology analysis signal regions after the maximum likelihood fit, for events in the high purity category. Upper row: 2L channel, electrons (left) and muons (right); middle row: 1L channel, electron (left) and muon (right); lower row: 0L channel. The plain red histograms represent the signal contribution normalized by the post-fit value of $\mu_{\vhcc}$, while the red line histograms show the expected signal contribution multiplied by a factor 100.}
\label{fig:sdmass-comp-2l}
\end{figure}

In the merged-jet topology analysis, the \vhcc\ signal is extracted via a binned maximum likelihood fit to the soft-drop mass \msd\ of \PH,
with the signal regions and the control regions from all three purity categories included in the fit simultaneously. In total, 15 bins are used in the fit for each region, with a bin width of 10\GeV corresponding roughly to the \msd\ resolution. The \msd\ distributions of the \vhcc\ and background processes in all three channels in the high-purity category are shown in Fig.~\ref{fig:sdmass-comp-2l}. The background prediction is in good agreement with the observed data, within uncertainties.

Similar to the resolved-jet topology analysis, the full procedure of the merged-jet topology analysis is validated by measuring the product of the \VZ production cross section and $\mathcal{B}\left(\PZ \to \cq\cqbar \right)$ normalised to the SM prediction. The event selection, including the kinematic BDT, the $\ccbar$ tagging discriminant criteria, and the signal extraction procedure, remain unchanged.
In place of \vhcc, the \vzcc\ process is considered to be the signal and \vhcc\ contributes to the background with cross section fixed to the SM prediction.
The measured signal strength is $\mu_{\vzcc}=0.69^{+0.89}_{-0.75}$ with an observed (expected) significance of 0.9 $(1.3)\,\sigma$. The results are consistent within uncertainties with the SM expectation.

The best fit value of $\mu$ for SM \vhcc\ production is $\mu_{\vhcc}=21^{+26}_{-24}$, and the observed (expected) UL on $\mu$ is found to be 71 ($49^{+24}_{-15}$) at 95\% \CL. The uncertainties in the expected UL correspond to a variation of $\pm1\,\sigma$ in the expected event yields under the background-only hypothesis. The observed values are in agreement with the SM expectation. The results for each channel and their combination are shown in Table~\ref{tab:limits-ind}. All channels yield comparable sensitivity in the merged-jet topology analysis.

\begin{table*}[htbp]
\topcaption{Observed and expected UL at 95\% CL on the signal strength $\mu$ for the \vhcc\ production for the resolved-jet and merged-jet topology analyses, which have a significant overlap. The results are also shown separately for each analysis channel.}
\label{tab:limits-ind}
\centering
{
\begin{tabular}{lcccccccc}\hline
& \multicolumn{4}{c}{Resolved-jet (inclusive)} & \multicolumn{4}{c}{Merged-jet (inclusive)} \\
& 0L & 1L & 2L & All channels & 0L & 1L & 2L & All channels\\  \hline \\
Expected UL  & $84_{-24}^{+35}$ & $79_{-23}^{+34}$ & $59_{-17}^{+25}$ & $38_{-11}^{+16}$ & $81_{-24}^{+39}$ & $88_{-27}^{+43}$ & $90_{-29}^{+48}$ & $49_{-15}^{+24}$ \\
Observed UL  & 66           & 120          & 116          & 75            & 74          & 120          & 76           & 71 \\ \hline
\end{tabular}
}
\end{table*}

\subsection{Combination}
\label{sec:combination}
To further improve the sensitivity of the search, a single likelihood analysis has been carried out on the two sets of data selected in the merged- and resolved-jet topology analyses.
To this end, the two analyses are categorised based on \ptV.
Events with values smaller than a certain value of \ptV are used in the resolved-jet topology analysis,
whereas the remaining events are used in the merged-jet topology analysis. The main theoretical and experimental systematic uncertainties
are correlated between the two analyses, with the exception of those related to the charm tagger efficiency (merged-jet topology)
and reshaping (resolved-jet topology), and those related to the \PV{}+jets PDFs and the renormalisation and factorisation scales because of the
different treatment of the \PV{}+jets processes adopted for the two analyses. The two regions demarcated by $\ptV=300\GeV$ provide
the best combined sensitivity in terms of expected limits on \vhcc.

The combination is validated by measuring the \vzcc\ signal strength. The measured value is $\mu_{\vzcc}=0.55^{+0.86}_{-0.84}$ with an observed (expected) significance of 0.7 $(1.3)\,\sigma$.

\begin{figure}[!hp]
\centering
\includegraphics[width=0.65\textwidth]{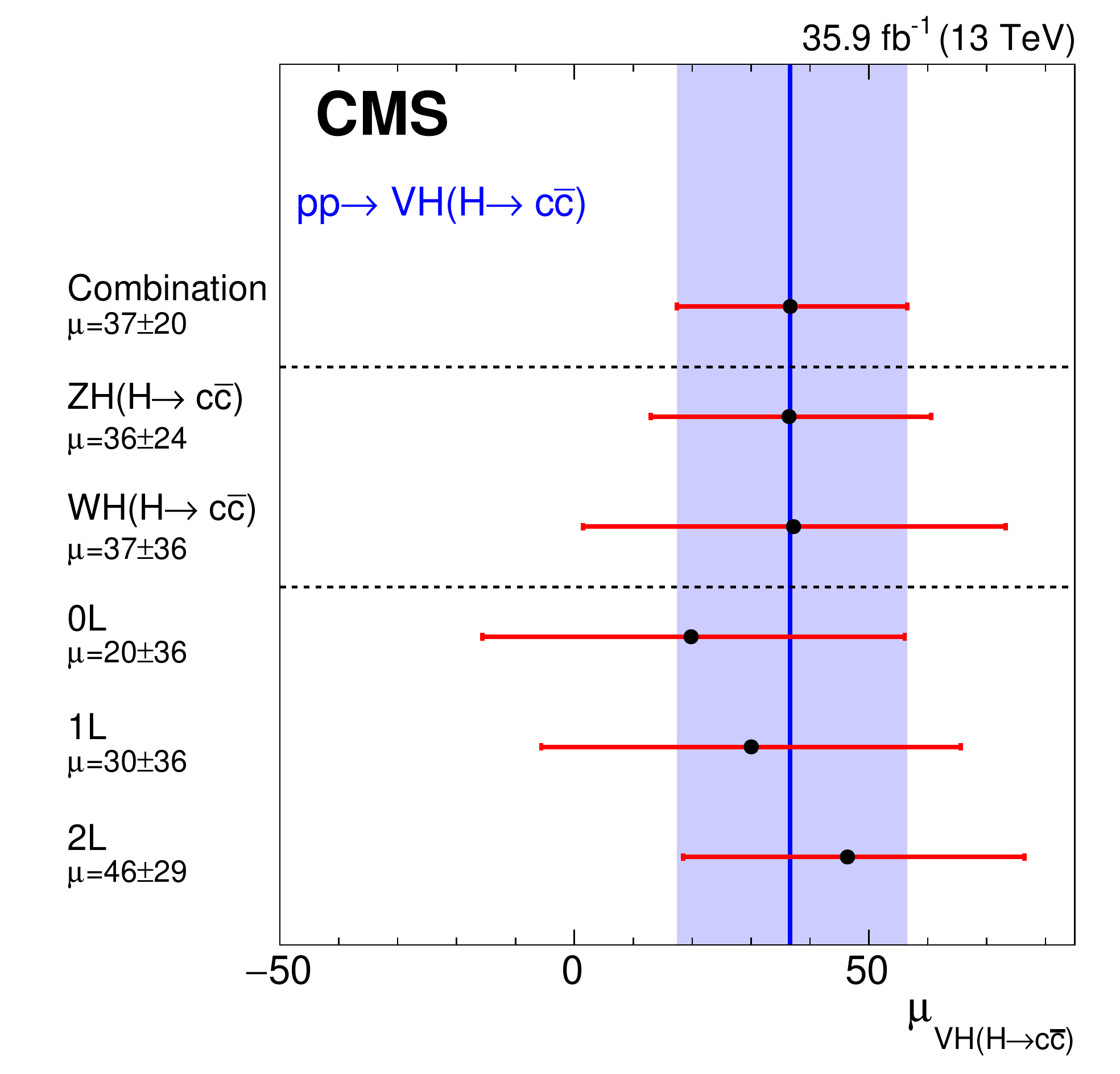}
\caption{\label{fig:comb-vhcc-plot-1}The fitted signal strength $\mu$ for the \zhcc and \whcc processes, and in each individual channel (0L, 1L, and 2L). The vertical blue line corresponds to the best fit value of $\mu$ for the combination of all channels, while the light-purple band corresponds to the $\pm1\,\sigma$ uncertainty in the measurement.}
\end{figure}

\begin{figure}[!hp]
\centering
\includegraphics[width=0.68\textwidth]{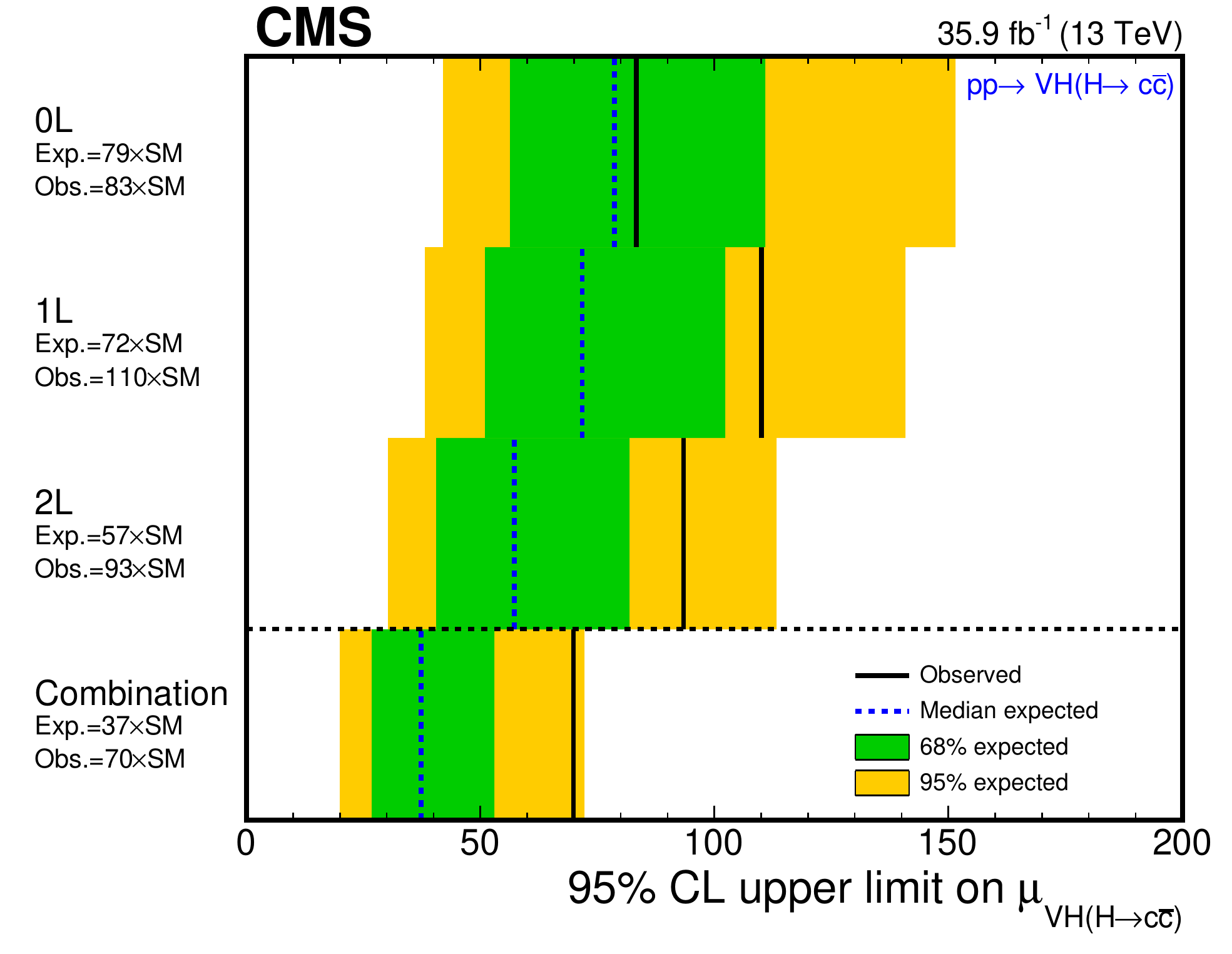}\\
\caption{\label{fig:comb-vhcc-plot-2}The 95\% CL upper limits on $\mu$ for the \vhcc\ process from the combination of the resolved-jet and merged-jet topology analyses in the different channels (0L, 1L, and 2L) and combined. The inner (green) and the outer (yellow) bands indicate the regions containing 68\% and 95\%, respectively, of the distribution of limits expected under the background-only hypothesis.}
\end{figure}

The fitted \vhcc\ signal strength from the combination of the two analyses after the selection on $\ptV$ is shown for all channels combined, per production process and per single analysis channel, in Fig.~\ref{fig:comb-vhcc-plot-1}.
The 95\% CL upper limits on the signal strength $\mu_{\vhcc}$ of each individual analysis after the selection on $\ptV$ and their combination is presented in Table~\ref{tab:comb-vhcc} and displayed in Fig.~\ref{fig:comb-vhcc-plot-2}.
The observed (expected) UL on $\sigma\left(\VH\right)\mathcal{B}\left(\PH \to \ccbar \right)$ obtained in the combined analysis is 4.5 $(2.4^{+1.0}_{-0.7})$\pb at 95\% \CL,
which is equivalent to an observed (expected) upper limit on $\mu$ of 70 ($37^{+16}_{-11}$) at 95\% confidence level.
The uncertainties in the expected UL correspond to a variation of $\pm1\,\sigma$ in the expected event yields under the background-only hypothesis. The measured signal strength is $\mu_{\vhcc}=37^{+17}_{-17}\stat^{+11}_{-9}\syst$. The observed values of the signal strength agrees within $2.1\,\sigma$ with the SM expectation. The results in the individual channels also agree with the SM expectation. The modest disagreement between the expected and observed UL's is mainly due to the small excess observed in data in the \Zee\ high-\ptV channel of the resolved-jet topology analysis.

\begin{table*}[!htbp]
\topcaption{\label{tab:comb-vhcc}The 95\% \CL upper limits on the signal strength $\mu_{\vhcc}$ for the \vhcc\ process, for the resolved-jet topology analysis for $\ptV<300\GeV$, the merged-jet topology analysis for $\ptV \geq 300\GeV$, and their combination.}
\centering
\begin{tabular}{lcccccc}
\hline
\multicolumn{7} {c} { 95\% \CL exclusion limit on $\mu_{\vhcc}$} \\
 & Resolved-jet & Merged-jet & \multicolumn{4}{c} {Combination} \\  \hline
 & ($\ptV<300\GeV$) & ($\ptV\geq300\GeV$) & 0L & 1L & 2L & All channels\\
Expected & $45_{-13}^{+18}$ & $73_{-22}^{+34}$ & $79_{-22}^{+32}$ & $72_{-21}^{+31}$  & $57_{-17}^{+25}$ & $37_{-11}^{+16}$ \\
Observed & 86 & 75 & 83 & 110 & 93 & 70 \\ \hline
\end{tabular}
\end{table*}

\newpage

\section{Summary}
In this paper, we present the first search by the CMS Collaboration for the standard model (SM) Higgs boson \PH decaying
to a pair of charm quarks, produced in association with a vector boson \PV\ (\PW or \PZ). The search uses proton-proton collision data at a
centre-of-mass energy of 13\TeV collected with the CMS detector in 2016 and corresponding to an integrated luminosity of 35.9\fbinv.
The search is carried out in five modes, \ZmmH, \ZeeH, \ZnnH, \WmnH, and \WenH,
with two complementary analyses targeting different regions of phase space. The signal is extracted by statistically
combining the results of the two analyses. Each analysis is first validated by carrying out a search for \PZ boson decay to a $\ccbar$ pair and comparing the observed signal strength with the SM prediction.
The \PZ boson signal strength for the combination of the two analyses is measured to be $\mu_{\vzcc}=\sigma/\sigma_\text{SM}=0.55^{+0.86}_{-0.84}$, with an observed (expected) significance of 0.7 (1.3) standard deviations.

The measured best fit value of $\sigma\left(\VH\right)\mathcal{B}\left(\PH \to \ccbar \right)$ for the combination of the two analyses
is $2.40^{+1.12}_{-1.11}\stat^{+0.65}_{-0.61}\syst$\pb, which corresponds to a best fit value of $\mu$ for SM \vhcc\ production of $\mu_{\vhcc}=\sigma/\sigma_\text{SM}=37^{+17}_{-17}\stat^{+11}_{-9}\syst$,
compatible within two standard deviations with the SM prediction. The larger measured $\mu_{\vhcc}$ value is due to a small excess observed in data in the resolved analysis, with a local significance of 2.1
standard deviations. The observed (expected) 95\% \CL upper limit on $\sigma\left(\VH\right)\mathcal{B}\left(\PH \to \ccbar \right)$ from the combination of the two analyses is 4.5 ($2.4^{+1.0}_{-0.7}$)\pb.
This limit can be translated into an observed (expected) upper limit on $\mu_{\vhcc}$ of 70 $(37^{+16}_{-11})$ at 95\% \CL by using the theoretical values of the cross section and branching fraction for the SM \PH boson with the mass of 125\GeV.
This result is the most stringent limit on $\sigma\left(\Pp\Pp\to \VH \right)\mathcal{B}\left(\HCC \right)$ to-date.

\begin{acknowledgments}
We congratulate our colleagues in the CERN accelerator departments for the excellent performance of the LHC and thank the technical and administrative staffs at CERN and at other CMS institutes for their contributions to the success of the CMS effort. In addition, we gratefully acknowledge the computing centres and personnel of the Worldwide LHC Computing Grid for delivering so effectively the computing infrastructure essential to our analyses. Finally, we acknowledge the enduring support for the construction and operation of the LHC and the CMS detector provided by the following funding agencies: BMBWF and FWF (Austria); FNRS and FWO (Belgium); CNPq, CAPES, FAPERJ, FAPERGS, and FAPESP (Brazil); MES (Bulgaria); CERN; CAS, MoST, and NSFC (China); COLCIENCIAS (Colombia); MSES and CSF (Croatia); RPF (Cyprus); SENESCYT (Ecuador); MoER, ERC IUT, PUT and ERDF (Estonia); Academy of Finland, MEC, and HIP (Finland); CEA and CNRS/IN2P3 (France); BMBF, DFG, and HGF (Germany); GSRT (Greece); NKFIA (Hungary); DAE and DST (India); IPM (Iran); SFI (Ireland); INFN (Italy); MSIP and NRF (Republic of Korea); MES (Latvia); LAS (Lithuania); MOE and UM (Malaysia); BUAP, CINVESTAV, CONACYT, LNS, SEP, and UASLP-FAI (Mexico); MOS (Montenegro); MBIE (New Zealand); PAEC (Pakistan); MSHE and NSC (Poland); FCT (Portugal); JINR (Dubna); MON, RosAtom, RAS, RFBR, and NRC KI (Russia); MESTD (Serbia); SEIDI, CPAN, PCTI, and FEDER (Spain); MOSTR (Sri Lanka); Swiss Funding Agencies (Switzerland); MST (Taipei); ThEPCenter, IPST, STAR, and NSTDA (Thailand); TUBITAK and TAEK (Turkey); NASU (Ukraine); STFC (United Kingdom); DOE and NSF (USA).

\hyphenation{Rachada-pisek} Individuals have received support from the Marie-Curie programme and the European Research Council and Horizon 2020 Grant, contract Nos.\ 675440, 752730, and 765710 (European Union); the Leventis Foundation; the A.P.\ Sloan Foundation; the Alexander von Humboldt Foundation; the Belgian Federal Science Policy Office; the Fonds pour la Formation \`a la Recherche dans l'Industrie et dans l'Agriculture (FRIA-Belgium); the Agentschap voor Innovatie door Wetenschap en Technologie (IWT-Belgium); the F.R.S.-FNRS and FWO (Belgium) under the ``Excellence of Science -- EOS" -- be.h project n.\ 30820817; the Beijing Municipal Science \& Technology Commission, No. Z181100004218003; the Ministry of Education, Youth and Sports (MEYS) of the Czech Republic; the Deutsche Forschungsgemeinschaft (DFG) under Germany’s Excellence Strategy -- EXC 2121 ``Quantum Universe" -- 390833306; the Lend\"ulet (``Momentum") Programme and the J\'anos Bolyai Research Scholarship of the Hungarian Academy of Sciences, the New National Excellence Program \'UNKP, the NKFIA research grants 123842, 123959, 124845, 124850, 125105, 128713, 128786, and 129058 (Hungary); the Council of Science and Industrial Research, India; the HOMING PLUS programme of the Foundation for Polish Science, cofinanced from European Union, Regional Development Fund, the Mobility Plus programme of the Ministry of Science and Higher Education, the National Science Center (Poland), contracts Harmonia 2014/14/M/ST2/00428, Opus 2014/13/B/ST2/02543, 2014/15/B/ST2/03998, and 2015/19/B/ST2/02861, Sonata-bis 2012/07/E/ST2/01406; the National Priorities Research Program by Qatar National Research Fund; the Ministry of Science and Education, grant no. 3.2989.2017 (Russia); the Programa Estatal de Fomento de la Investigaci{\'o}n Cient{\'i}fica y T{\'e}cnica de Excelencia Mar\'{\i}a de Maeztu, grant MDM-2015-0509 and the Programa Severo Ochoa del Principado de Asturias; the Thalis and Aristeia programmes cofinanced by EU-ESF and the Greek NSRF; the Rachadapisek Sompot Fund for Postdoctoral Fellowship, Chulalongkorn University and the Chulalongkorn Academic into Its 2nd Century Project Advancement Project (Thailand); the Nvidia Corporation; the Welch Foundation, contract C-1845; and the Weston Havens Foundation (USA).
\end{acknowledgments}

\bibliography{auto_generated}
\cleardoublepage \appendix\section{The CMS Collaboration \label{app:collab}}\begin{sloppypar}\hyphenpenalty=5000\widowpenalty=500\clubpenalty=5000\vskip\cmsinstskip
\textbf{Yerevan Physics Institute, Yerevan, Armenia}\\*[0pt]
A.M.~Sirunyan$^{\textrm{\dag}}$, A.~Tumasyan
\vskip\cmsinstskip
\textbf{Institut f\"{u}r Hochenergiephysik, Wien, Austria}\\*[0pt]
W.~Adam, F.~Ambrogi, T.~Bergauer, M.~Dragicevic, J.~Er\"{o}, A.~Escalante~Del~Valle, M.~Flechl, R.~Fr\"{u}hwirth\cmsAuthorMark{1}, M.~Jeitler\cmsAuthorMark{1}, N.~Krammer, I.~Kr\"{a}tschmer, D.~Liko, T.~Madlener, I.~Mikulec, N.~Rad, J.~Schieck\cmsAuthorMark{1}, R.~Sch\"{o}fbeck, M.~Spanring, D.~Spitzbart, W.~Waltenberger, C.-E.~Wulz\cmsAuthorMark{1}, M.~Zarucki
% ...
\vskip\cmsinstskip
\dag: Deceased\\
1:  Also at Vienna University of Technology, Vienna, Austria\\
% ...
\end{sloppypar}
\end{document}